# СОВРЕМЕННАЯ И МЕЛ–КАЙНОЗОЙСКАЯ ДИВЕРСИФИКАЦИЯ ПОКРЫТОСЕМЕННЫХ

## Шереметьев С.Н., Чеботарева К.Е.

## Ботанический институт им. В.Л.Комарова РАН
## 197376 С.–Петербург, ул. Проф. Попова, 2

*E–mail: sn.sheremetiev@gmail.com*

Мел-кайнозойская история покрытосеменных растений имела результатом появление определенного характера распределения числа таксонов разного уровня (числа видов и родов в семействах, отношения S/G в семействах, числа видов в родах). В большинстве случаев эти распределения удовлетворительно описываются степенным законом (распределение Парето). В логарифмической системе координат степенная функция имеет вид прямой линии. Эмпирические кривые достаточно хорошо повторяют эту линию, однако в правых частях графиков, т.е. в областях малых объемов таксонов, наблюдается заметное отклонение эмпирических кривых от теоретических. Это свидетельствует о том, для полного соответствия теоретическим кривым объемы малых таксонов должны быть значительно больше. Моделирование соотношений числа родов и видов в семействах показало, что только в случае использования динамического фактора вымирания наблюдается удовлетворительное соответствие наблюдаемого и расчетного числа видов в широком диапазоне итераций. Это позволило предположить, что в ходе эволюции покрытосеменных растений имело место дифференцированное вымирание видов. Это подразумевает, что в родах с большим числом видов темпы вымирания должны были быть минимальными. Напротив, при уменьшении числа видов – коэффициенты вымирания могли увеличиваться на порядки. В результате большие роды становились еще больше, а малые – уменьшались. При этом частотное распределение видов в родах изменялось по степенному закону. Первоначальное расхождение численности таксонов, которое обусловило их дальнейшее разделение на большие и малые, могло быть вызвано появлением и экспансией трав с их функциональными и адаптивными возможностями.

Ключевые слова: покрытосеменные, семейства, роды, диверсификация

Диверсификация, то есть процесс и результат изменения биологического разнообразия, широко освещается в мировой литературе с разных точек зрения. Диверсификация покрытосеменных растений интенсивно изучается в последние десятилетия не только на морфологическом, но и на молекулярном уровне, что нашло отражение в создании и развитии филогенетической системы (APG, 1998; APG II, 2003; APG III, 2009; Chase, Reveal, 2009; Haston et al., 2009; Reveal, Chase, 2011 и др., APG IV, 2016). Оставляя в стороне такие аспекты биоразнообразия, как биогеографический, экосистемный, ценотический, сосредоточимся только на одном плане диверсификации, а именно таксономическом. В свою очередь таксономическое разнообразие мы попытаемся описать только с количественной стороны, опираясь на филогенетические построения APG III. В первой части мы разберем статистические свойства этой системы, затем попытаемся приблизиться к решению вопроса о соотношении числа родов и видов в семействах покрытосеменных (что может быть причиной наблюдаемых соотношений в настоящее время и в прошлом), рассмотрим изменение числа родов и семейств в мелу–кайнозое, а также динамику скорости диверсификации в этом интервале времени.

## МАТЕРИАЛ И МЕТОДИКА

Информация об объеме семейств и родов (т.е. о количестве родов в каждом семействе и видов в каждом роде) была получена на ресурсе 'The Plant List'[1]. Полный список семейств, родов и видов, информация о которых приведена на этом сайте, можно найти в файле 'AngiospermsWholeTaxonList.xlsx'[2]. Для 9 семейств, которые не включены в 'The Plant List', информация об объеме таксонов приводится по P.F. Stevens (2001). В расчеты принимались только признанные (accepted) по версии 'The Plant List' таксоны.

Преобладающие жизненные формы (ЖФ) определены для 2022 родов покрытосеменных (см. электронное Приложение[3]). Для этого были использованы ресурсы, перечисленные в табл. 1. В некоторых монотипных семействах жизненные формы могли быть определены по L. Watson, M.J. Dallwitz (1992).

Встречались следующие сочетания жизненных форм внутри родов:

| | | |
|---|---|---|
| A | Arborescent | |
| C | Cactaceae | |
| H | Herbs | |
| L | Lianas | |
| P | Palms | |
| S | Shrubs | |
| T | Trees | |
| HA | Herbs, Arborescent | |
| HL | Herbs, Lianas | |
| HS | Herbs, Shrubs | |
| HSA | Herbs, Shrubs, Arborescent | |
| HSL | Herbs, Shrubs, Lianas | |
| HSLT | Herbs, Shrubs, Lianas, Trees | |
| HST | Herbs, Shrubs, Trees | |
| LT | Lianas, Trees | |
| SA | Shrubs, Arborescent | |
| SL | Shrubs, Lianas | |
| SLT | Shrubs, Lianas, Trees | |
| ST | Shrubs, Trees | |

---

[1] http://www.theplantlist.org/1.1/browse/A/
[2] https://www.dropbox.com/s/b8fw6iym0fpd5df/AngiospermsWholeTaxonList.xlsx?dl=0
[3] https://www.dropbox.com/s/39u0475yajfa6ko/Diversification–Supplement.xlsx?dl=0

Для неформальных групп системы APG III (см. рис. 1) был рассчитан индекс травянистости ($I_H$). Этот индекс представляет собой отношение суммы встречаемости ЖФ трав к сумме встречаемости других жизненных форм (во всех их сочетаниях) для всех родов группы:

$$I_H = \frac{\sum H}{\sum C + \sum L + \sum P + \sum S + \sum T}$$

В этой формуле буквенные обозначения соответствуют обозначениям, приведенным выше в списке сочетаний жизненных форм. Предположим, например, что в группе, состоящей из родов X, Y и Z, наблюдается такое сочетание жизненных форм: X – ST; Y – HSL; Z – H. Тогда искомый индекс равен 2/4 (2H и L2ST). Если этот индекс меньше единицы и приближается к нулю, то в группе преобладают роды иных жизненных форм, чем травы. Когда $I_H$ больше единицы – в группе преобладают травы.

Первое появление родов в палеонтологической летописи было определено по Paleobiology Database[4] (PBDB). Был просмотрен весь список родов, представленный в The Plant List. Для 804 таксонов из этого списка удалось найти их возраст (соответствующие списки семейств и родов, а также необходимые ссылки см. в электронном Приложении). Первое появление 249 семейств покрытосеменных в палеонтологической летописи зафиксировано в PBDB и ряде работ (Brown, 1962; Daghlian, 1981; Muller, 1981; Zavada, Benson, 1987; Krutzsch, 1989; Taylor, 1990; Collinson et al., 1993; Conran et al., 1994; Herendeen, Crane, 1995; Wing, Boucher, 1998; Poole, Francis, 1999; Doyle, 2000; Lee et al., 2001; Crepet et al., 2004; Friis et al., 2004, 2011; Riley, Stockey, 2004; Zhi–Chen et al., 2004; Prasad et al., 2005; Sille et al., 2006; Stockey, 2006; Ramírez et al., 2007; Smith, Stockey, 2007a; Smith, Stockey, 2007b; Zhang et al., 2007; Smith et al., 2008; Yamada et al., 2008; Conran et al., 2009; Smith et al., 2009; Taylor et al., 2009; Manchester, O'Leary, 2010; Martínez–Millán, 2010; Carvalho et al., 2011; Daly et al., 2011; Grímsson et al., 2011; Muellner, 2011; Prasad et al., 2011; Birch et al., 2012; Palazzesi et al., 2012; Collinson et al., 2012; Coiffard et al., 2013a; Coiffard et al., 2013b; Mendes et al., 2014). Этот ряд датировок мы назвали Fossil-ряд.

Возраст семейств оценивался также по результатам молекулярного датирования (Wikström et al., 2001; Bremer et al., 2004; Janssen, Bremer, 2004; Renner, 2005; Biffin et al., 2010; Ocampo, Columbus, 2010; Xi et al., 2012; Mennes et al., 2013; Magallón et al., 2015) (см. электронное Приложение). Были построены два ряда датировок – Wikström–ряд (для

---

[4] http://www.paleobiodb.org/cgi–bin/bridge.pl?a=displaySearchColls&type=view

345 семейств) и Magallón–ряд (для 363 семейств) (названные по преобладающему вкладу соответствующих авторов). Все оценки приведены для стволовых групп (stem groups) (см. Павлинов, 2005).

Абсолютную скорость диверсификации для стволовых групп можно оценить по формуле (Magallón, Castillo, 2009):

$$\hat{r}_\varepsilon = \frac{\ln[n(1-\varepsilon) + \varepsilon]}{t}, \qquad (1)$$

где $t$, в данном случае, время от возникновения клады до настоящего времени; $n$ – число появившихся видов; $\varepsilon$ – относительная скорость вымирания, которая определяется отношением скорости вымирания к скорости видообразования. Поскольку эти скорости чаще всего неизвестны, то $\varepsilon$ можно принимать равной числам в диапазоне от 0 (отсутствие вымираний) до 0.9 (максимально возможные скорости, поскольку при равенстве скоростей вымирания и видообразования никакой диверсификации в указанный промежуток времени наблюдаться не будет). Ниже мы покажем результаты моделирования диверсификации, которые позволяют приближенно оценить скорости вымирания.

Для того чтобы иметь возможность оценить динамику процесса диверсификации в каждой из эпох мела–кайнозоя мы несколько модифицировали формулу (1):

$$\hat{r}_i = \frac{n_i(1-\varepsilon) + \varepsilon}{t_i}, \qquad (2)$$

где $n_i$ – число таксонов, появившихся в эпоху $i$ (отсчет ведется от раннего мела к интервалу времени, в котором мы объединили плиоцен, плейстоцен и голоцен), $t_i$ – продолжительность этой эпохи.

Для определения относительной скорости диверсификации мы использовали следующее выражение:

$$r_i = \frac{(n_{i+1} - n_i)(1-\varepsilon) + \varepsilon}{t_i - t_{i+1}} = \frac{\Delta n(1-\varepsilon) + \varepsilon}{\Delta t}, \qquad (3)$$

где $\Delta t$ – разность (в млн. лет) между серединами соседних эпох $i$ и $i+1$; $\Delta n$ – разность числа таксонов, появившихся в эпохах $i+1$ и $i$. Отсюда видно, что положительные значения $r_i$ могут наблюдаться только в том случае, если в более позднюю эпоху появилось больше таксонов, чем в предыдущую. Иначе скорости диверсификации будут отрицательными. Пока неизвестна достаточно ясная картина появления и вымирания видов в эпохи мела-кайнозоя, формулы (2, 3) могут быть использованы для оценки $\hat{r}_i$ и $r_i$ для ро-дов и семейств. Поскольку датировки Fossil-ряда были получены только для ныне живущих родов, то в этом случае $\varepsilon = 0$. Для семейств также $\varepsilon = 0$, так как практически неизвестны вымершие таксоны этого ранга (Collinson et al., 1993).

Данные по размерам геномов покрытосеменных были получены с ресурса «C–values database» (Bennett, Leitch, 2012).

Доверительные интервалы (P=0.05) (независимые от формы распределения) для квантилей оценивали по формулам (Bland, 2000):

$$j = nq - 1.96\sqrt{nq(1-q)}$$
$$k = nq + 1.96\sqrt{nq(1-q)},$$

где *j* и *k* – номера (округляются до следующего целого) членов ранжированного вариационного ряда *x*; *n* – объем ряда; *q* – порядок квантили (например, квантиль порядка 0.1 или нижняя дециль, квантиль порядка 0.25 или нижняя квартиль, квантиль порядка 0.5 или медиана и т.д.). Таким образом, нижняя граница доверительного интервала равна *x(j)*, а верхняя – *x(k)*.

Коэффициенты детерминации, приведенные в работе, статистически значимы на доверительном уровне 0.05 (если не указано иное).

# СОВРЕМЕННАЯ ДИВЕРСИФИКАЦИЯ

## Общие вопросы

Согласно классификации наземных растений, сопровождающей систему APG III (Chase, Reveal,2009), покрытосеменные (Magnoliidae) являются подклассом класса Equisetopsida. В этот класс также включаются подклассы Anthocerotidae, Bryidae, Marchantiidae, Lycopodiidae, Equisetidae, Marattiidae, Ophioglossidae, Polypodiidae, Psilotidae, Ginkgooidae, Cycadidae, Pinidae, Gnetidae (Chase, Reveal,2009).

В подклассе Magnoliidae содержится 18 надпорядков, 68 порядков, 414 семейств (Reveal, Chase, 2011). Необходимо отметить, что эти числа неокончательные. Например, в APG IV выделены 64 порядка и 416 семейство (APG IV, 2016).

По версии 'The Plant List' (с небольшими дополнениями по Stevens, 2001) в семействах покрытосеменных признано 14583 рода и 304351 вид.

## Ряды распределения таксонов

Первым, кто заинтересовался объемом таксонов применительно к некоторым аспектам биогеографии и эволюции, был, судя по всему, Ч. Дарвин (Darwin, 1872; см. также его письма Д. Хукеру за 1844–1858 гг.: Darwin Correspondence Database). Затем J.C.

Willis (Willis, 1922, 1940; Willis, Yule, 1922) на основе данных своей предыдущей книги (Willis, 1919) количественно показал характер распределения родов покрытосеменных в зависимости от количества входящих в них видов. Позже G.U. Yule (1925), используя эти материалы, теоретически обосновал открытое J.C. Willis распределение, которое известно, как распределение Юла (Reed, Hughes, 2007). G.U. Yule (1925) показал, как должна изменяться численность таксонов во времени. Эта работа получила развитие в ряде публикаций в недавнее время (Reed, Hughes, 2002, 2007). Эта тема также рассматривается в ряде современных работ (Поздняков, 2005; Minelli et al., 1991; Chu, Adami, 1999; Hilu, 2006; Mora et al., 2011; Strand, Panova, 2015 и др.)

Ряд распределения числа видов в каждом из родов покрытосеменных (рис. 2а, б), построенный по данным ресурса 'The Plant List', демонстрирует ярко выраженную асимметрию. Это распределение, очевидно, является негауссовым и принадлежит семейству гиперболических распределений, однако нам не удалось подобрать соответствующий закон (на статистически значимом уровне). Это отличается от результатов, показанных J.C. Willis (Willis, 1922, 1940). Наши попытки воспроизвести эти результаты по исходным материалам (Willis, 1919) также окончились неудачей. Воспроизведенные графики больше походили на наши (рис. 2а, б), чем на оригинальные рисунки J.C. Willis.

Распределение число видов можно представить несколько по–иному. После группировки родов в семействах рассчитывается отношение числа видов к числу родов каждого семейства (S/G ratio). Распределение этого ряда чисел (рис. 2в, г) удовлетворительно описывается степенным законом (распределение Парето; о свойствах этого распределения см.: Кох, 2002). Это же относится к рядам распределений числа видов (рис. 2д, е) и родов (рис. 2ж, з) в семействах. Общим свойством этих эмпирических распределений является заметное отклонение от теоретических кривых в правых частях графиков (рис. 2б, г, е, ж), т.е. в областях малых объемов таксонов. Для полного соответствия теоретическим кривым объемы этих таксонов должны быть значительно больше. Ниже мы попытаемся показать (в виде предположения, следующего из простой численной модели), что это может быть связано преимущественно с одним фактором, а именно с дифференцированным вымиранием в таксонах разного объема. В больших таксонах скорости вымирания должны быть на порядки меньше, чем в малых. Относительно высокие скорости вымирания в небольших таксонах и обусловливают, судя по всему, отмеченное отклонение. Несколько перефразируя одно из популярных описаний закона Парето можно сказать, что в результате дифференцированного вымирания большие таксоны становятся еще больше, а малые – уменьшаются.

## Соотношение числа родов и видов в семействах

Соотношение числа родов и видов в семействах исследовалось многими авторами (Wilkinson, 1999; Bertrand et al., 2006; Balmford et al., 2006a, 2006b; Palmer et al., 2008; Kallimanis et al., 2012; Qian, Zhang, 2014 и др.). Преимущественно это были данные по различным локальным и региональным флорам и фаунам.

Мы попытались проанализировать указанное соотношение в таксономическом плане. Количество родов тесно коррелирует с количеством видов в семействах покрытосеменных независимо от того, данные какой системы были использованы: APG III (рис. 3а) ('The Plant List'), А. Энглера (рис. 3б) (Willis, 1919) или А.Л. Тахтаджяна (рис. 3в) (Takhtajan, 2009).

Для голосеменных (рис. 3г) такая корреляция выражена значительно слабее. Это может быть следствием значительных вымираний в этой группе растений. По системе С.В. Мейена (1992) из 41 семейства голосеменных вымерла большая часть, а именно 29 семейств.

Кроме того, можно легко показать, что корреляция между числом родов и видов в разных группах растений определенным образом зависит от объема семейств. Так, если мы возьмем только те семейства, число родов в которых меньше 10, то во всех случаях такая корреляция окажется статистически не значимой (рис. 4). Однако, начиная с 20 родов в семействах (для покрытосеменных – с 30), коэффициенты детерминации быстро увеличиваются и всегда статистически значимы. Когда объем семейств достигает 70 и более родов коррелированность числа родов и видов уже не снижается. Поэтому можно предположить, что такое соотношение для голосеменных (рис. 3г) отражает в какой–то степени их не очень удачную, хотя и очень продолжительную, эволюционную историю.

В семействах птеридофитов (рис. 3д) и бриофитов (рис. 3е) корреляция между числом родов и видов выражена вполне отчетливо и почти не уступает таковой у покрытосеменных растений (рис. 3а–в).

Такая же корреляция наблюдается в разных группах животных: млекопитающих (рис. 5а), рептилий (рис. 5б), амфибий (рис. 5в), птиц (рис. 5г), рыб (рис. 5д), пауков (рис. 5е).

Наличие тесных корреляций между числом родов и видов в семействах самых разных групп растений и животных (рис. 4 и 5) заставляет предположить, что соответствующие системы вполне адекватно отражают реальные объемы родов и семейств. Это под-тверждается и хорошим соотношением объемов таксонов в этих группах (рис. 6). Помимо этого, проблема больших родов (см. Frodin, 2004) может получить более или менее приемлемое объяснение.

Для того чтобы попробовать приблизиться к решению этого вопроса мы написали программу (на VBA for Excel) численного моделирования процесса дивергенции. Эта программа очень проста и не претендует на сколько-нибудь всеобъемлющее описание этого явления. Нас интересовал только один вопрос: можно ли, изменяя одно или два условия, получить результаты, достаточно близкие к реальным? Первые варианты программы рассчитывали количество родов и видов в семействах. При небольшом числе итераций (до 5–7) время вычислений было большим, но более или менее приемлемым. Когда число итераций превысило 10, стало понятно, что необходимые вычисления могут продлиться не один месяц в течение круглых суток. Поэтому было принято решение проводить вычисления без расчета числа родов, но основываясь на уже имеющихся данных.

В программе изменялись только два показателя: число итераций и коэффициент (точнее индекс – см. ниже) вымирания. Для индекса вымирания задавалась только верхняя граница, а конкретное его значение в заданных границах определял генератор случайных чисел. То же самое относится и к числу дочерних видов (параметру дивергенции).

Для расчетов было удобно разделить коэффициент вымирания $k$ на индекс и фактор:
$$k = E/F,$$
где $E$ – индекс вымирания (задается пользователем), $F$ – фактор вымирания (может изменяться динамически, в зависимости от числа видов, образовавшихся на предыдущем шаге, либо явно задаваться пользователем) (см. табл. 2). Тогда коэффициенты вымирания рассчитываются следующим образом:
$$k_i = \frac{1}{F} Rand(0, B_E),$$
где $i$ – номер итерации; $Rand$ – оператор генератора случайных чисел, выдающего целочисленные значения в границах, указанных в скобках; $B_E$ – верхняя граница индекса вымирания (задается пользователем).

На первом шаге количество видов полагается следующим:
$$n_1 = Int\big([Rand(0, D)] \cdot (1 - k_1)\big),$$
где $n_1$ – число видов, которое рассчитывается в первой итерации; $Int$ – оператор округления до следующего целого; $D$ – верхняя граница числа дочерних видов.

Количество видов в следующих итерациях:

$$n_i = Int\{(n_{i-1} + [Rand(0, D)]) \cdot (1 - k_i)\},$$

где $n_i$ – число видов, которое рассчитывается на текущем шаге $i$.

Эти вычисления ($k_i$ и $n_i$, где $i$ изменяется от 1 до заданного числа итераций) повторяются для каждого семейства.

Можно, вслед за Ч. Дарвином (Darwin, 1872, p. 91), предположить, что между итерациями или актами видообразования проходят тысячи поколений. Весьма вероятно, что каждая форма остается неизменной в течение длительных периодов времени, а затем вновь подвергается модификациям (Darwin, 1872, p. 91). Вместе с тем, чем больше таких актов, тем выше индексы вымирания, необходимые для достижения близких к реальным результатов (табл. 2).

Примеры вычислений при разных условиях выполнения программы (рис. 7) показывают, что наилучшие и стабильные результаты появляются только тогда, когда используется динамический фактор вымирания, который зависит от числа видов, рассчитанных на предыдущих шагах (его значения показаны в табл. 2). Самым главным свойством фактора вымирания является то, что он уменьшает коэффициент вымирания при увеличении числа видов. При этом коэффициент изменяется не в разы, а на порядки. Большое число рассчитанных вариантов хорошо соответствует реальным данным только при задании такого динамического фактора вымирания (рис. 7а–в). Это соответствие стабильно сохраняется на протяжении 14–60 итераций (рис. 8а–в). Нужно отметить, что число итераций и индекс вымирания оказались очень тесно взаимосвязанными (рис. 8г). Если мы увеличивали число итераций, то было необходимо увеличивать и индекс вымирания.

В случае использования статического фактора вымирания в подавляющем большинстве случаев результаты были неудовлетворительными (рис. 7г, д). Только однажды удалось получить удовлетворительную сходимость реальных и расчетных данных (рис. 7е, табл. 2). В дальнейшем при увеличении числа итераций такая сходимость резко уменьшалась (рис. 8в).

В результате мы сделали предположение, что в ходе эволюции покрытосеменных растений имело место дифференцированное вымирание видов. В родах с большим числом видов темпы вымирания должны были быть минимальными. Напротив, при уменьшении числа видов – коэффициенты вымирания могли увеличиваться на порядки. В результате большие роды становились еще больше, а малые – уменьшались. При этом частотное распределение видов в родах изменялось по степенному закону (рис. 2).

Помимо этого, очень важно отметить то, что коэффициенты вымирания, которые приводят к наблюдаемой численности таксонов, очень малы и вряд ли когда-нибудь превышали 0.1 (см. формулы 1 и 2). Данные палеонтологии также свидетельствуют о том, что скорости видообразования значительно превышали скорости вымирания (Friis et al., 2011; см. также De Vos, 2015). Проблема больших родов находит, возможно, свое решение. Роды с большим числом видов являются не таксономическими артефактами (см. Strand, Panova, 2015), а следствием дифференцированных скоростей вымирания в таксонах разного объема и отражают, судя по всему, естественное положение дел. Однако остается открытым вопрос о причине первоначального расхождения численности таксонов.

**Возможные причины расхождения численности таксонов**

Одним из основных отличительных свойств таксонов являются жизненные формы растений, присущие им в разных сочетаниях. Мы сгруппировали роды, в которых имеются растения только одной жизненной формы, по этому признаку. Объемы таксонов разного ранга строго коррелированы между собой (рис. 9а, б, в). Показательно, что наибольший вклад в таксономическое разнообразие вносят травы. С большим отставанием от них следуют деревья и кустарники, представляющие, тем не менее, большое количество семейств, родов и видов. Эти основные жизненные формы отличаются и разными размерами геномов (2С) (рис. 9г). Травы значительно превосходят деревья и кустарники по этому признаку (см. также Шереметьев и др., 2011). Возможно, увеличение 2С является одной из необходимых предпосылок увеличения потенциала биологического разнообразия.

В связи с этим нужно рассмотреть вопрос о том, как изменяется 2С у групп разного систематического положения (табл. 3). Судя по всему, связь размеров геномов с положением групп разного уровня в системе APG III (пока) обнаружить нельзя. Это касается как крупных групп (табл. 3, верхняя часть), так и клад нижнего уровня (не имеющих в своем составе других вложенных неформальных групп) (табл. 3, нижняя часть) (отметим, что лучшими оценками центральных тенденций соответствующих рядов исходных данных из-за характера их распределения являются медианы, а не средние). Также отсутствует корреляция между 2С и объемами таксонов разного ранга в этих группах. Вместе с тем, индекс травянистости, рассчитанный для клад нижнего уровня, тесно коррелирует с размером генома (рис. 10). Это означает, что по отношению к размеру генома имеет значение не само по себе систематическое положение группы, но ее состав жизненных форм. Раз-мер геном увеличивается по мере увеличения вклада трав в таксономическое разнообразие. В этом ряду есть только одно исключение – Commelinids. Высокий индекс травянистости этой группы по непонятным причинам не вполне сочетается с указанной тенденцией (рис. 10).

В литературе обсуждаются два варианта возможных последствий увеличения генома – рост функционального потенциала и расширение адаптивных возможностей вида (Grime et al., 1985; Gregory, 2001, 2002; Vinogradov, 2001, 2003). Первое подтверждается более интенсивным функционированием трав по сравнению с кустарниками и деревьями. Обработка базы данных GLOPNET (Wright et al., 2004) показала, что интенсивность дыхания, а также максимальная фотосинтетическая способность растений (Слемнев, 1988) (потенциальный фотосинтез, Amax) значительно выше у трав, чем у кустарников и деревьев (рис. 11).

Рост содержания ДНК может рассматриваться в качестве одного из эффективных средств адаптогенеза. Например, отмечен на порядок больший геном и значительно более широкий экологический диапазон трав неогена относительно древесных форм палеогена (Шереметьев и др., 2011). Травы и кустарники менее специализированы, чем деревья. Они способны обитать в тех климатических условиях, где деревьям нет места. Травяные биомы вытеснили в неогене лесные на значительных территориях по причине большей эвритопности их представителей (Шереметьев, Гамалей, 2009). Кроме того, участие этих жизненных форм в составе родов покрытосеменных зеркально изменялось в мелу–кайнозое – участие трав увеличивалось, а деревьев – уменьшалось (рис. 12а). Особенно заметно это происходило, начиная с олигоцена. Одной из возможных причин такого явления могло быть уменьшение концентрации углекислого газа в атмосфере (рис. 12б). К олигоцену относят, например, возникновение $C_4$–трав, появившихся в ответ на изменение состава атмосферы (уменьшение концентрации $CO_2$). В миоцене начался процесс широкой экспансии травяных биомов и постепенное вытеснение ими биомов с преобладанием древесной растительности (Шереметьев, Гамалей, 2009).

Таким образом, одной из возможных причин первоначального расхождения численности таксонов могли быть появление и экспансия трав, с их функциональными и адаптивными возможностями. Однажды появившиеся различия в численности таксонов, судя по всему, поддерживались разными скоростями вымирания, в результате чего большие таксоны становились еще больше, а малые – уменьшались.

# МЕЛ–КАЙНОЗОЙСКАЯ ДИВЕРСИФИКАЦИЯ

Мел-кайнозойскую диверсификацию покрытосеменных растений можно наглядно представить (по меньшей мере) в виде четырех кривых: 1) кумулятивной кривой, показывающей сумму числа таксонов, появившихся в текущую (т.е. не в современную, а в ту, которая рассматривается в текущий момент) и предшествующие эпохи; 2) кривой известных изменений числа таксонов в каждую из эпох; 3) кривой абсолютных скоростей диверсификации (формула 2), необходимость расчета которых вызвана различной продолжительностью эпох мела-кайнозоя; 4) кривой относительных скоростей диверсификации (формула 3), демонстрирующей изменение числа новых таксонов в текущую эпоху относительно предыдущей. Рассмотрим эти кривые отдельно для родов и семейств покрытосеменных (напомним, что все данные приведены для ныне существующих таксонов, т.е. в обоих случаях $\varepsilon = 0$).

## Роды

Суммарное количество родов покрытосеменных растений увеличивалось во все эпохи мела-кайнозоя (рис. 13а). В общем случае (для всех родов) кумулятивная кривая описывается (с наилучшим приближением) гиперболой (рис. 13б). Это означает, что пока не обнаружено снижение темпов диверсификации на уровне родов и кривая не выходит на уровень равновесия (на плато). Этот случай не является уникальным. Разнообразие фанерозойской морской биоты также аппроксимируется гиперболой (Марков, Коротаев, 2007; в этой работе можно подробно ознакомиться с историей вопроса). Такой рост биоразнообразия может быть обусловлен внутренними закономерностями (положительными обратными связями второго порядка: Марков, Коротаев, 2007), а также реакцией на давление климата. Это может относиться как к экспоненциальным моделям (Hewzulla et al., 1999), так и к сигмоидным до достижения ими фазы равновесия.

Динамика разнообразия родов, представленных жизненной формой 'деревья', с наилучшим приближением описывается одной из сигмоидных кривых ('модель Ратковского') (рис. 13в). Однако визуально пока не заметно, что кривая выходит на плато, хотя, судя по всему, и стремится к состоянию равновесия. Роды травянистых растений, напротив, демонстрируют ярко выраженный экспоненциальный рост разнообразия (рис. 13г). Абсолютные скорости диверсификации (рис. 13д) всех трех групп растений также хорошо описываются экспоненциальными кривыми (рис. 13е-з).

Эти изменения разнообразия покрытосеменных растений происходили на фоне значительных климатических перестроек мела-кайнозоя. На протяжении этого времени наблюдается неуклонное снижение концентрации углекислого газа в атмосфере (рис. 14а). Реконструкция климатической зональности всех эпох мела на основе литологических, палеонтологических и геохимических индикаторов климата показала, что в нижнем мелу между 30° с.ш. и 40°-50° ю.ш. располагался эвапоритовый аридный пояс (Жарков и др., 2004; Чумаков, 2004а). Тропический гумидный пояс начал появляться только в альбе и постепенно расширялся до конца мела (Чумаков, 2004б). Это несколько расходится с другими палеоклиматическими реконструкциями (Boucot et al., 2013), в которых, однако, также отмечено минимальное развитие влажного тропического климата в раннем мелу (рис. 14б). Пояса наибольшей продуктивности растений располагались в средних и высоких широтах (Чумаков, 2004б).

В мелу и до конца эоцена планета находилась в режиме «теплой биосферы» (Чумаков, 2004а, б) или «warm mode» («greenhouse mode») (Frakes et al., 1992). Полностью отсутствовали ледяные шапки на полюсах, а температурный градиент «полюс-экватор» мог составлять 0.1°C на 1° широты (против современного значения 0.6°C на 1° широты) (Amiot, 2004; см. также Nikolaev et al., 1998). Однако вектор снижения глобальной температуры (несмотря на некоторые колебания) отчетливо проявился на протяжении позднего мела-кайнозоя (рис. 14а). Оледенение Антарктиды, начавшееся на границе эоцена и олигоцена, означало переход Земли в состояние "холодной биосферы" (Ахметьев, 2004), продолжающееся до нашего времени. Это сопровождалось значительным уменьшением концентрации углекислого газа в атмосфере, снижением осадков, аридизацией, сокращением поясов умеренного климата (рис. 14) на фоне значительного понижения уровня моря (Miller et al., 2005).

Климатические перестройки кайнозоя явно не благоприятствовали процветанию растений, однако вопреки этому абсолютная скорость диверсификации покрытосеменных увеличивалась (рис. 14а). Это могло быть связано со значительным увеличением (примерно в 6 раз) температурного градиента 'полюс-экватор', что должно означать не просто похолодание климата в средних и высоких широтах, а усиление дифференциации экотопов, увеличение их количества и разнообразия. Кроме того, нельзя забывать о внутренней обусловленности динамики разнообразия (Hewzulla et al., 1999; Марков, Коротаев, 2007), которая ведет процесс в определенном направлении (рис. 13), несмотря на давление среды.

Это давление хорошо видно при исследовании относительной скорости диверсификации (рис. 15). В общем случае (для всех родов) заметно небольшое снижение этой скорости в палеоцене, значительное ее уменьшение в олигоцене и в последние 5 млн. лет (в плиоцене, плейстоцене, голоцене) (рис. 15а). Это может быть связано с расширением аридных поясов в соответствующие эпохи (рис. 14б, 15б). Сходным образом на аридизацию климата реагируют роды древесных растений (рис. 15в). У родов, представленных травами, относительная скорость диверсификации, возможно, увеличивается по мере расширения поясов умеренного прохладного климата (рис. 15в). Здесь следует отметить, что возникновение открытых травяных экосистем относят к олигоцену, а начало их широкой экспансии, развитие собственно травяных биомов – к миоцену (Sheremet'ev, Gamalei, 2012). Этот факт, судя по всему, находит отражение в резком увеличении относительной скорости диверсификации трав после миоцена (рис. 15а).

Таким образом, именно роды трав обеспечили эволюционный прорыв покрытосеменных в неогене и четвертичном периоде (рис. 9, 13д, 15а). Возможно, этот прорыв был связан с большим размером генома (рис. 9г, 10), более интенсивным функционированием (рис. 11), несравненно более коротким жизненным циклом, по сравнению с другими жизненными формами. Более того, начиная с миоцена, травяные биомы начали вытеснять лесные на больших пространствах суши в силу большей экологической подвижности и эвритопности (Шереметьев, Гамалей, 2009). В этом эволюционном движении неогена-антропогена участвовали, конечно, не все роды, а лишь небольшая их часть (10-15%) (рис. 2). Подавляющее большинство родов покрытосеменных сохраняли эволюционное равновесие, поддерживая достигнутый уровень разнообразия.

### Семейства

Выше было отмечено, что в этом исследовании имеется три ряда датировок возраста семейств (Fossil, Magallón и Wikström). Заранее нельзя отдать предпочтение ни одному из них, поэтому мы проведем анализ со всеми этими массивами данных.

Кумулятивные кривые изменения разнообразия семейств (рис. 16а) наилучшим образом аппроксимируются сигмоидными кривыми (рис. 16б-г). Это свидетельствует об уменьшении скорости роста разнообразия семейств (в отличие от родов) и приближении их числа к некоторому максимальному уровню и постепенном выходе на плато. Такое поведение демонстрируют все три ряда, но особенно показательным в этом смысле является Magallón-ряд (рис. 16в). Отсюда можно предположить, что эволюция покрытосеменных растений будет продолжатся (если не учитывать антропогенный стресс), но в основном за счет изменения разнообразия родов, а не семейств.

В этом отношении весьма показательны кривые изменения числа семейств, впервые появившихся в эпохи мела-кайнозоя (рис. 17а). Эти кривые практически одинаковы по форме для всех трех рядов датировок. Наибольшее число семейств появилось в позднем мелу. Второй пик разнообразия, хотя и значительно меньшего масштаба, приходится на эоцен. Однако абсолютные скорости диверсификации (число семейств, отнесенное к продолжительности эпохи) (рис. 17б) заметно отличаются от этого (рис. 17а), хотя и указывают на то, что наибольшие скорости образования семейств приходились на поздний мел, палеоцен и эоцен (Magallón- и Wikström-ряды). По данным палеонтологической летописи абсолютная скорость диверсификации покрытосеменных в кайнозое изменялась не столь значительно (рис. 17б).

Относительные скорости диверсификации (рис. 17в) были минимальными в палеоцене и олигоцене, максимальными – в позднем мелу, эоцене и миоцене. Последнее хорошо соответствует климатическим оптимумам мела-кайнозоя (см. Шереметьев, Гамалей, 2009). Если олигоценовый минимум диверсификации совпадает с переходом Земли в режим «холодной биосферы», то палеоценовый – с некоторым понижением глобальной температуры в палеоцене (рис. 17в). Нельзя исключить, что это было следствием катастрофических событий на рубеже мела-палеогена (Renne et al., 2015 и др.). Другие корреляции относительной скорости диверсификации с климатическими изменениями мела-кайнозоя не прослеживаются.

Количество семейств, появившихся в эпохи мела-кайнозоя хорошо соотносится с развитием площадей аридных территорий (рис. 18а-в). Увеличение этих площадей сопровождается уменьшением количества семейств в соответствующие эпохи. Оценки этих корреляций статистически значимы для всех рядов датировок возраста семейств. Абсолютные скорости диверсификации статистически значимо коррелируют с глобальными континентальными осадками (рис. 18д, е) и уровнем моря (рис. 18з, и) для Magallón- и Wikström-рядов. Слабые тенденции такого рода (не подтверждаемые статистически) наблюдаются и для Fossil-ряда (рис. 18г, ж). Таким образом, нельзя исключить, что активность глобального гидрологического цикла влияет на изменение динамики разнообразия семейств покрытосеменных растений. В эпохи с низкой активностью этого цикла (усиление аридизации климата) скорости диверсификации снижаются. Наоборот, высокий уровень моря и большое количество континентальных осадков могут способствовать появлению большего числа семейств покрытосеменных растений.

## ЗАКЛЮЧЕНИЕ

Мел-кайнозойская история покрытосеменных растений имела результатом появление определенного характера распределения числа таксонов разного уровня (числа видов и родов в семействах, отношения S/G в семействах, числа видов в родах). В большинстве случаев эти распределения удовлетворительно описываются степенным законом (распределение Парето). В логарифмической системе координат степенная функция имеет вид прямой линии. Эмпирические кривые достаточно хорошо повторяют эту линию, однако в правых частях графиков, т.е. в областях малых объемов таксонов, наблюдается заметное отклонение эмпирических кривых от теоретических. Это свидетельствует о том, для полного соответствия теоретическим кривым объемы малых таксонов должны быть значительно больше. Моделирование соотношений числа родов и видов в семействах показало, что только в случае использования динамического фактора вымирания наблюдается удовлетворительное соответствие наблюдаемого и расчетного числа видов в широком диапазоне итераций. Это позволило предположить, что в ходе эволюции покрытосеменных растений имело место дифференцированное вымирание видов. Это подразумевает, что в родах с большим числом видов темпы вымирания должны были быть минимальными. Напротив, при уменьшении числа видов – коэффициенты вымирания могли увеличиваться на порядки. В результате большие роды становились еще больше, а малые – уменьшались. При этом частотное распределение видов в родах изменялось по степенному закону. Первоначальное расхождение численности таксонов, которое обусловило их дальнейшее разделение на большие и малые, могло быть вызвано появлением и экспансией трав с их функциональными и адаптивными возможностями.

Абсолютные скорости диверсификации родов покрытосеменных растений наилучшим образом аппроксимируются экспонентами. Наибольшие темпы роста разнообразия наблюдаются в неогене. Относительные скорости диверсификации достаточно хорошо отражают экологические аспекты динамики разнообразия. Роды древесных растений отзываются на усиление аридизации климата снижением темпов диверсификации. Роды травянистых растений относительная скорость диверсификации увеличивается по мере расширения поясов умеренного прохладного климата.

Кумулятивные кривые изменения разнообразия семейств наилучшим образом аппроксимируются сигмоидными кривыми. Это свидетельствует об уменьшении скорости роста разнообразия семейств (в отличие от родов) и приближении их числа к некоторому максимальному уровню и постепенном выходе на плато. Абсолютные скорости диверсификации статистически значимо коррелируют с глобальными континентальными осадками и уровнем моря. Поэтому нельзя исключить, что активность глобального гидрологического цикла влияла на изменение динамики разнообразия семейств покрытосеменных растений.

## БЛАГОДАРНОСТИ

## СПИСОК ЛИТЕРАТУРЫ

# ПОДПИСИ К РИСУНКАМ
# («СОВРЕМЕННАЯ И МЕЛ–КАЙНОЗОЙСКАЯ ДИВЕРСИФИКАЦИЯ ПОКРЫТОСЕМЕННЫХ»)

**Рис. 1.** Соотношения порядков и некоторых семейств покрытосеменных в системе APG III (по: APG III, 2009, с изменениями). Справа – сокращения для названий неформальных групп системы (ANITA – ANITA grade, Mag – Magnoliids, blMag – basal lineage of Magnoliids, Com – Commelinids, nComM – non Commelinids Monocots, blE – basal lineages of Eudicots, blCE – basal lineages of Core Eudicots, Fab – Fabids, Mal – Malvids, blR – basal lineage of Rosids, blR – basal lineages of Asterids, Lam – Lamiids, Cam – Campanulids).

**Рис. 2.** Распределения числа видов в родах (а, б), отношения числа видов к числу родов (S/G) в семействах (в, г), числа видов (д, е) и числа родов в семействах (ж, з) покрытосеменных растений. Ряды чисел таксонов сортировались по убыванию и максимальным значениям присваивался ранг равный единице. Справа (б, г, е, з) эти распределение показаны в логарифмическом масштабе. Прямые линии (б, г, е, з) – степенные функции, которыми аппроксимировались исходные данные (а, в, д, ж) и уравнения для них.

**Рис. 3.** Соотношение числа родов и видов в семействах покрытосеменных (а, б, в), голосеменных (г), птеридофитов (д) и бриофитов (е). По данным 'The Plant List' (а, г, д, е), J.C. Willis (1919; система А. Энглера) (б), А.Л. Тахтаджяна (Takhtajan, 2009) (в).

**Рис. 4.** Рис. 4. Коррелированность числа родов и видов в семействах покрытосеменных, птеридофитов и бриофитов в зависимости от объема семейств (объяснения в тексте). $r^2$-кр. – кривая критических значений, ниже которой коэффициенты детерминации статистически не значимы.

**Рис. 5.** Соотношение числа родов и видов в семействах млекопитающих (Wilson, Reeder, 2005) (а), рептилий (Uetz, 2010) (б), амфибий (AmphibiaWeb, 2015) (в), птиц (Bock, Farrand, 1980) (г), рыб (Нельсон, 2009) (д), пауков (Jocqué, Dippenaar-Schoeman, 2006) (е).

**Рис. 6.** Рис. 6. Соотношения объемов таксонов разных рангов в группах растений и животных. 1 – покрытосеменные, 2 – голосеменные, 3 – птеридофиты, 4 – бриофиты, 5 – млекопитающие, 6 – амфибии, 7 – рептилии, 8 – пауки, 9 – рыбы, 10 – птицы. Оси абсцисс и ординат показаны в логарифмическом масштабе.

**Рис. 7.** Соотношение реального и расчетного числа видов по результатам работы программа 'Divergence 3.0' (условия выполнения программы для этого рисунка даны в табл. 2).

**Рис. 8.** Анализ результатов работы программы 'Divergence 3.0'. а, б – зависимость максимального расчетного числа видов от числа итераций (а) и индексов вымирания (б); в – зависимость коррелированности между реальным и расчетным числом видов от число итераций в случаях статического и динамического факторов вымирания; г – соотношение между числом итераций и индексами вымирания.

**Рис. 9.** Соотношение объемов таксонов разных рангов после группировки родов покрытосеменных по жизненным формам (а, б, в; оси ординат в логарифмическом масштабе); размер генома (медианы, пикограммы) у разных жизненных форм покрытосеменных (г, вертикальные планки погрешностей – доверительные интервалы на уровне значимости 0.05). H – травы, S – кустарники, T – деревья, P – пальмы, L – лианы, C – представители сем. Cactaceae.

**Рис. 10.** Соотношение размера генома (медианы) и индекса травянистости ($I_H$) для неформальных групп системы APG III (см. рис. 1) (1 – ANITA grade, 2 – Magnoliids, 3 – basal lineages of Magnoliids, 5 – non Commelinids Monocots, 6 – basal lineages of Eudicots, 7 – basal lineages of Core Eudicots, 8 – Fabids, 9 – Malvids, 10 – basal lineages of Rosids, 11 – basal lineages of Asterids, 12 – Lamiids, 13 – Campanulids). Точка для Commelinids (Com) не включена в расчеты коэффициента детерминации (с этой точкой $r^2 = 0.43$, что статистически отличается от нуля на уровне значимости 0.05).

**Рис. 11.** Темновое дыхание (а) и максимальная реализация фотосинтетической способности (Amax) (б) (медианы, вертикальные планки погрешностей – доверительные интервалы на уровне значимости 0.05) у разных жизненных форм по результатам обработки базы данных GLOPNET (Wright et al., 2004).

**Рис. 12.** Участие жизненных форм (ЖФ) в родах покрытосеменных, появившихся впервые в палеонтологической летописи в эпохи мела-кайнозоя (а) и на фоне изменений в это же время относительного (к доиндустриальному уровню) содержания углекислого газа в атмосфере (Berner, 2006) (б).

**Рис. 13.** Изменения числа родов покрытосеменных, появившихся в палеонтологической летописи в эпохи мела-кайнозоя: а – кумулятивные кривые; б, в, г – аппроксимация кумулятивных кривых гиперболой (б, все роды), сигмоидной кривой (в, деревья), экспонентой (г, травы); д – абсолютные скорости диверсификации; е, ж, з – аппроксимация этих скоростей (е – все роды, ж – деревья, з – травы) экспонентами. Светлые сплошные кривые – границы доверительных интервалов, светлые прерывистые кривые – границы интервалов предсказания. Оси ординат (а, д) слева – для всех родов, справа – для деревьев и трав. 1– ранний мел; 2 – поздний мел; 3 – палеоцен; 4 – эоцен; 5 – олигоцен; 6 – миоцен; 7 – плиоцен, плейстоцен, голоцен.

**Рис. 14.** Климатические условия мела-кайнозоя. а – изменения абсолютной скорости диверсификации (СД) покрытосеменных на фоне климатических изменений мела-кайнозоя: континентальных осадков (Gibbs et al., 1999); глобальной температуры морской воды, рассчитанной по соотношению изотопов кислорода в раковинах бентосных и планктонных фораминифер (по формуле Lear et al., 2000) по результатам обработки базы данных J. Veizer et al. (1999); периодов частичного (штриховка) и полного (заливка) оледенения северного (1) и южного (2) полюсов (по: Zachos et al., 2001); относительного содержания углекислого газа в атмосфере ($RCO_2$) по данным модели GEOCARB III (Berner, Kothavala, 2001), континентальных осадков (Gibbs et al., 1999). б – площади территорий, занятых климатическими поясами (в процентах от общей площади суши в соответствующую эпоху) (по картам Boucot et al., 2013; береговые линии определены в программе GPlates ver. 1.5). Cr1 (Ber–Apt) – ранний мел (берриас–апт); Cr1 (Alb) – ранний мел (альб); Cr2 – поздний мел; Pa – палеоцен; Eo – эоцен; Ol – олигоцен; Mi – миоцен, PPH – плиоцен, плейстоцен, голоцен.

**Рис. 15.** Относительные скорости диверсификации ($\varepsilon = 0$) (а) и их соотношение для всех родов (б), деревьев (в) и трав (г) с площадями (по картам Boucot et al., 2013) аридных (б, в) и умеренно прохладных (г) территорий в эпохи мела-кайнозоя. Цифровые обозначения те же, что и на рис. 13.

**Рис. 16.** Изменения числа семейств покрытосеменных, появившихся в палеонтологической летописи в эпохи мела-кайнозоя: а – кумулятивные кривые; б, в, г – аппроксимация кумулятивных кривых (б – Fossil-ряд, в – Magallón-ряд, г – Wikström-ряд) сигмоидными функциями. Светлые сплошные кривые – границы доверительных интервалов; светлые прерывистые кривые – границы интервалов предсказания. Цифровые обозначения те же, что и на рис. 13.

**Рис. 17.** Изменения числа семейств покрытосеменных, появившихся в палеонтологической летописи в эпохи мела-кайнозоя: а – количество новых семейств, зафиксированных в каждую из эпох; б – то же количество, отнесенное к продолжительности эпох; в – относительная скорость диверсификации семейств покрытосеменных ($\varepsilon = 0$); T – глобальная температура морской воды (по данным Veizer et al., 1999; Lear et al., 2000) (усредненные по эпохам данные, отклонения от тренда). Цифровые обозначения те же, что и на рис. 13.

**Рис. 18.** Соотношение количества семейств, появившихся в эпохи мела-кайнозоя, с площадями аридных территорий (по картам Boucot et al., 2013): а – Fossil-ряд, б – Magallón-ряд, в – Wikström-ряд. Соотношение абсолютных скоростей диверсификации семейств покрытосеменных в эпохи мела-кайнозоя с континентальными осадками (Gibbs et al., 1999) (г, д, е) и уровнем моря (относительно современного) (Miller et al., 2005) (ж, з, и): г, ж – Fossil-ряд; д, з – Magallón-ряд; е, и – Wikström-ряд. Цифровые обозначения те же, что и на рис. 13.

ТАБЛИЦЫ

Таблица 1. Перечень ресурсов, использованных для определения жизненных форм родов

| Название ресурса | Интернет-ссылка |
|---|---|
| Flora of Chile | http://efloras.org/index.aspx |
| Flora of China | http://efloras.org/index.aspx |
| Flora of Missouri | http://efloras.org/index.aspx |
| Flora of Mozambique | http://www.mozambiqueflora.com/index.php |
| Flora of North America | http://efloras.org/index.aspx |
| Flora of West Tropical Africa | https://archive.org/download/FloraOfWestTropi00hutc/FloraOfWestTropicalAfrica-JohnHutchinson.pdf |
| Flora of Zimbabwe | http://www.zimbabweflora.co.zw/index.php |
| Flora Zambesiaca | http://apps.kew.org/efloras/search.do |
| Gateway to African Plants | http://gateway.myspecies.info/ |
| Jepson eFlora | http://ucjeps.berkeley.edu/IJM.html |
| Legumes of the world | http://www.kew.org/science-conservation/research-data/resources/legumes-of-the-world |
| List of Species of the Brazilian Flora | http://floradobrasil.jbrj.gov.br/jabot/listaBrasil/PrincipalUC/PrincipalUC.do?lingua=en |
| The Online World Grass Flora | http://www.kew.org/data/grasses-db.html |
| Trees and shrubs of the Andes of Ecuador | http://efloras.org/index.aspx |

Таблица 2. Параметры и условия выполнения программы Divergence 3.0 (для рисунка 7)

| Параметр | Расшифровка | Рисунок 7 | | | | | |
|---|---|---|---|---|---|---|---|
| | | а | б | в | г | д | е |
| It | Число итераций | 14 | 30 | 60 | 14 | 14 | 14 |
| D | Верхняя граница числа дочерних видов. Изменяется от 0 до D с помощью генератора случайных чисел. | 2 | 2 | 2 | 2 | 2 | 2 |
| $B_E$ | Верхняя граница индекса вымирания. Изменяется от 0 до $B_E$ с помощью генератора случайных чисел. | 4 | 9 | 20 | 3 | 3 | 3 |
| F | Фактор вымирания. | Динамический | | | $10^3$ | $10^4$ | $10^5$ |
| Динамический фактор вымирания | | | | | | | |
| | Число видов | $<10^2$ | $10^2-10^3$ | $10^3-10^4$ | $10^4-10^5$ | $>10^5$ | |
| | Фактор вымирания | $10^3$ | $10^4$ | $10^5$ | $10^6$ | $10^7$ | |

Таблица 3. Размеры геномов в группах системы APG III (см. рис. 1)

| Статистики | Angiosperms | BA | Mon | ED | CED | Ros | Ast |
|---|---|---|---|---|---|---|---|
| X | 12.4 | 4.7 | 22.7 | 6.0 | 5.2 | 3.8 | 5.7 |
| CI | 0.4 | 0.8 | 0.8 | 0.2 | 0.2 | 0.2 | 0.2 |
| Xb | 12.0 | 3.9 | 21.8 | 5.7 | 5.0 | 3.6 | 5.5 |
| Xu | 12.8 | 5.5 | 23.5 | 6.2 | 5.5 | 4.0 | 5.9 |
| n | 10683 | 111 | 4107 | 6464 | 6081 | 2962 | 2554 |
| Sx | 20.1 | 4.3 | 27.1 | 9.6 | 8.7 | 4.9 | 5.7 |
| q0.5b | 4.9 | 2.3 | 12.1 | 2.9 | 2.7 | 1.8 | 3.6 |
| q0.5 | 5.1 | 2.7 | 12.9 | 2.9 | 2.8 | 1.9 | 3.8 |
| q0.5u | 5.3 | 3.4 | 13.7 | 3.0 | 2.9 | 2.0 | 4.0 |

| Статистики | ANITA | Mag | blMag | Com | nComM | blE | blCE | Fab | Mal | blR | blA | Lam | Cam |
|---|---|---|---|---|---|---|---|---|---|---|---|---|---|
| X | 6.6 | 4.0 | 7.2 | 10.9 | 33.1 | 17.6 | 10.6 | 4.5 | 2.2 | 1.2 | 4.0 | 3.8 | 7.4 |
| CI | 2.4 | 0.8 | 1.6 | 0.5 | 1.4 | 1.5 | 1.8 | 0.2 | 0.2 | 0.4 | 0.7 | 0.3 | 0.3 |
| Xb | 4.2 | 3.2 | 5.6 | 10.4 | 31.8 | 16.2 | 8.7 | 4.2 | 2.0 | 0.8 | 3.3 | 3.5 | 7.1 |
| Xu | 9.0 | 4.8 | 8.9 | 11.4 | 34.5 | 19.1 | 12.4 | 4.7 | 2.3 | 1.6 | 4.6 | 4.0 | 7.8 |
| n | 24 | 84 | 3 | 1928 | 2179 | 383 | 565 | 2132 | 802 | 28 | 111 | 1092 | 1351 |
| Sx | 6.0 | 3.6 | 1.5 | 11.1 | 32.3 | 14.5 | 22.3 | 5.4 | 2.2 | 1.0 | 3.6 | 4.5 | 6.1 |
| q0.5b | 1.9 | 2.3 | N/A | 7.1 | 24.0 | 12.2 | 3.1 | 2.1 | 1.3 | 1.0 | 2.7 | 2.3 | 5.6 |
| q0.5 | 4.4 | 2.5 | 7.2 | 7.8 | 25.4 | 13.5 | 3.3 | 2.2 | 1.4 | 1.0 | 3.3 | 2.4 | 5.94 |
| q0.5u | 6.7 | 2.9 | 8.7 | 8.4 | 26.4 | 15.9 | 4.1 | 2.3 | 1.4 | 1.1 | 3.7 | 2.6 | 6.22 |

BA – basal angiosperms (ANITA + Mag + blMag), Mon – monocots, ED – eudicots, CED – core eudicots, Ros – rosids, Ast – asterids. Остальные сокращения см. в подписи к рис. 1. X – средние; CI – доверительные интервалы для средних; n – число определений 2C; Sx – стандартные отклонения; q0.5 – квантили порядка 0.5 (медианы); литеры 'b' и 'u' на последних позициях имен статистик обозначают соответственно нижние и верхние границы доверительных интервалов (на уровне значимости 0.05).

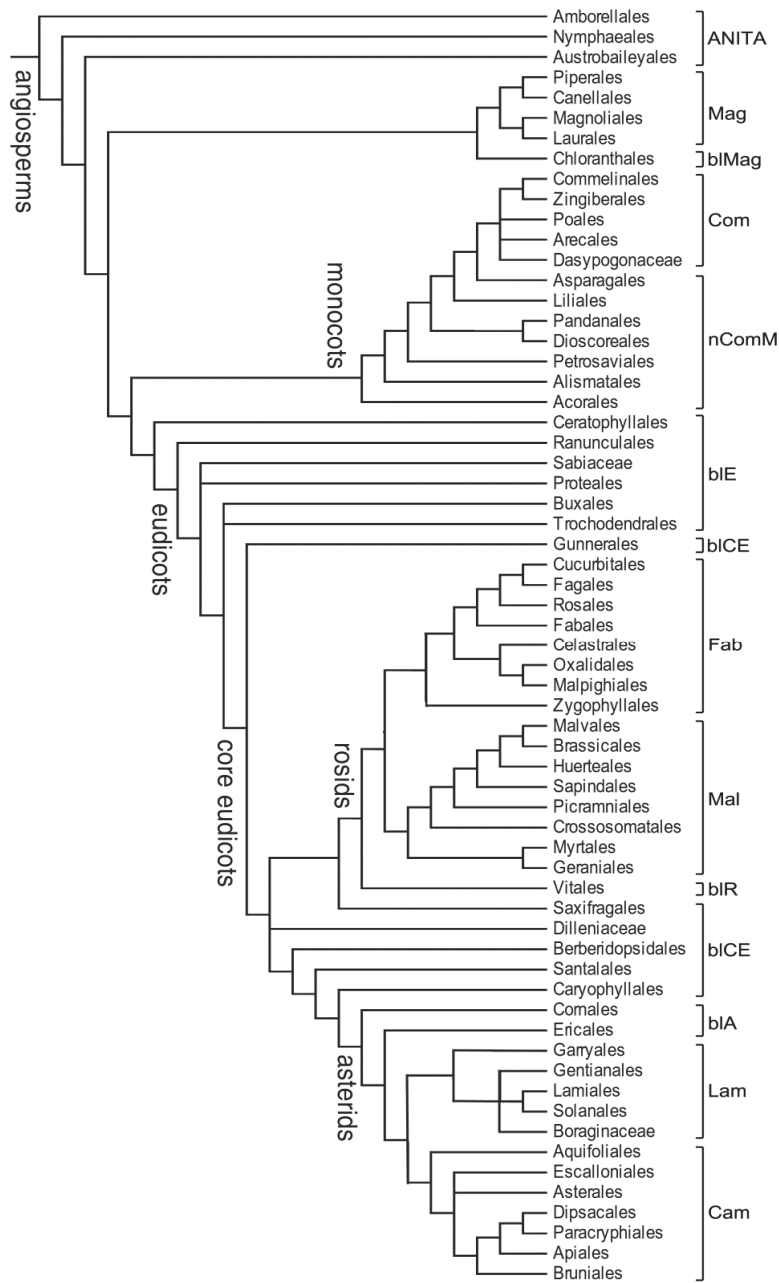

Рис. 1. Соотношения порядков и некоторых семейств покрытосеменных в системе APG III (по: APG III, 2009, с изменениями). Справа - сокращения для названий неформальных групп системы (ANITA - ANITA grade, Mag - Magnoliids, blMag - basal lineage of Magnoliids, Com - Commelinids, nComM - non Commelinids Monocots, blE - basal lineages of Eudicots, blCE - basal lineages of Core Eudicots, Fab - Fabids, Mal - Malvids, blR - basal lineage of Rosids, blR - basal lineages of Asterids, Lam - Lamiids, Cam - Campanulids).

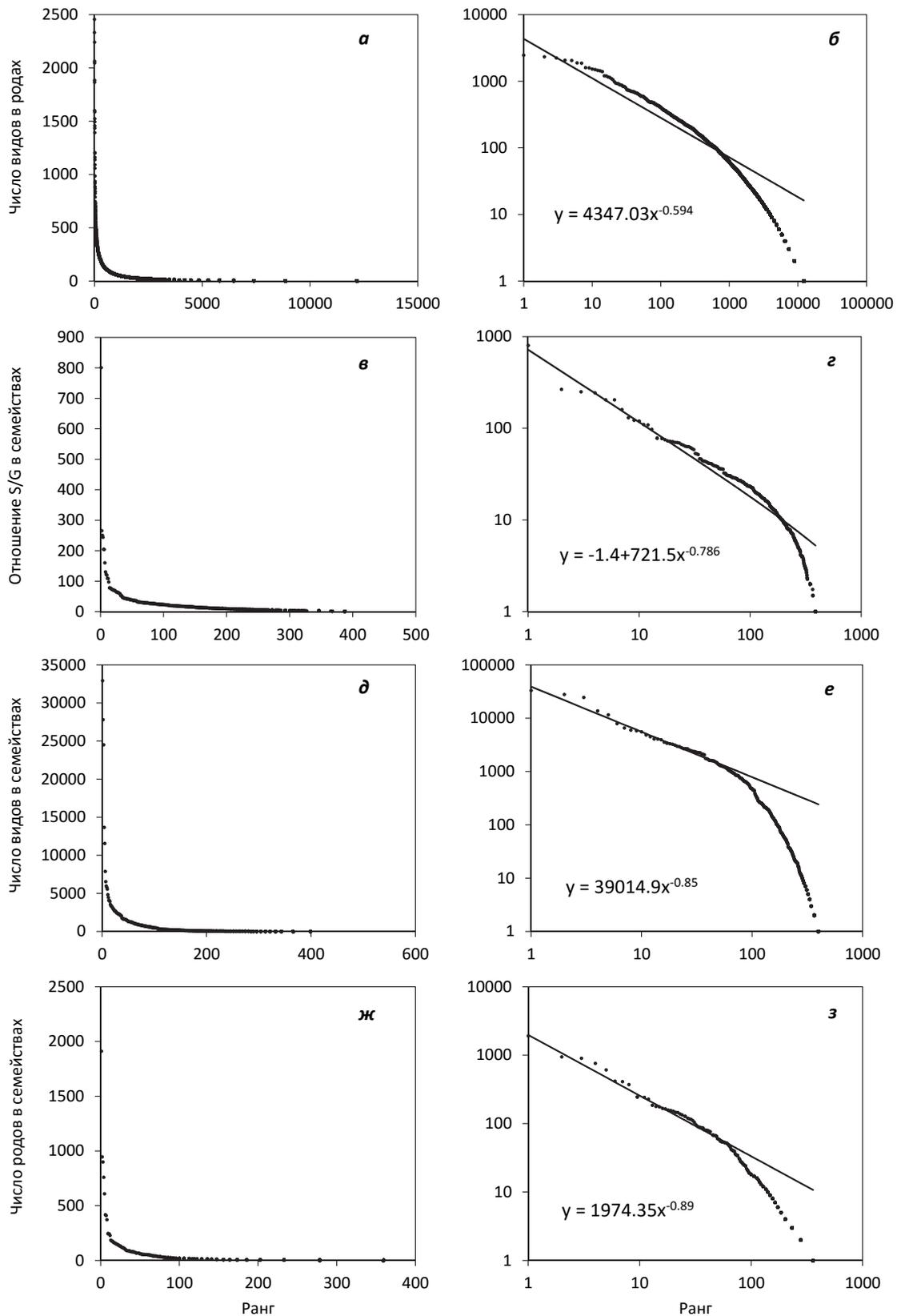

Рис. 2. Распределения числа видов в родах (а, б), отношения числа видов к числу родов (S/G) в семействах (в, г), числа видов (д, е) и числа родов в семействах (ж, з) покрытосеменных растений. Ряды чисел таксонов сортировались по убыванию и максимальным значениям присваивался ранг равный единице. Справа (б, г, е, з) эти распределение показаны в логарифмическом масштабе. Прямые линии (б, г, е, з) - степенные функции, которыми аппроксимировались исходные данные (а, в, д, ж) и уравнения для них.

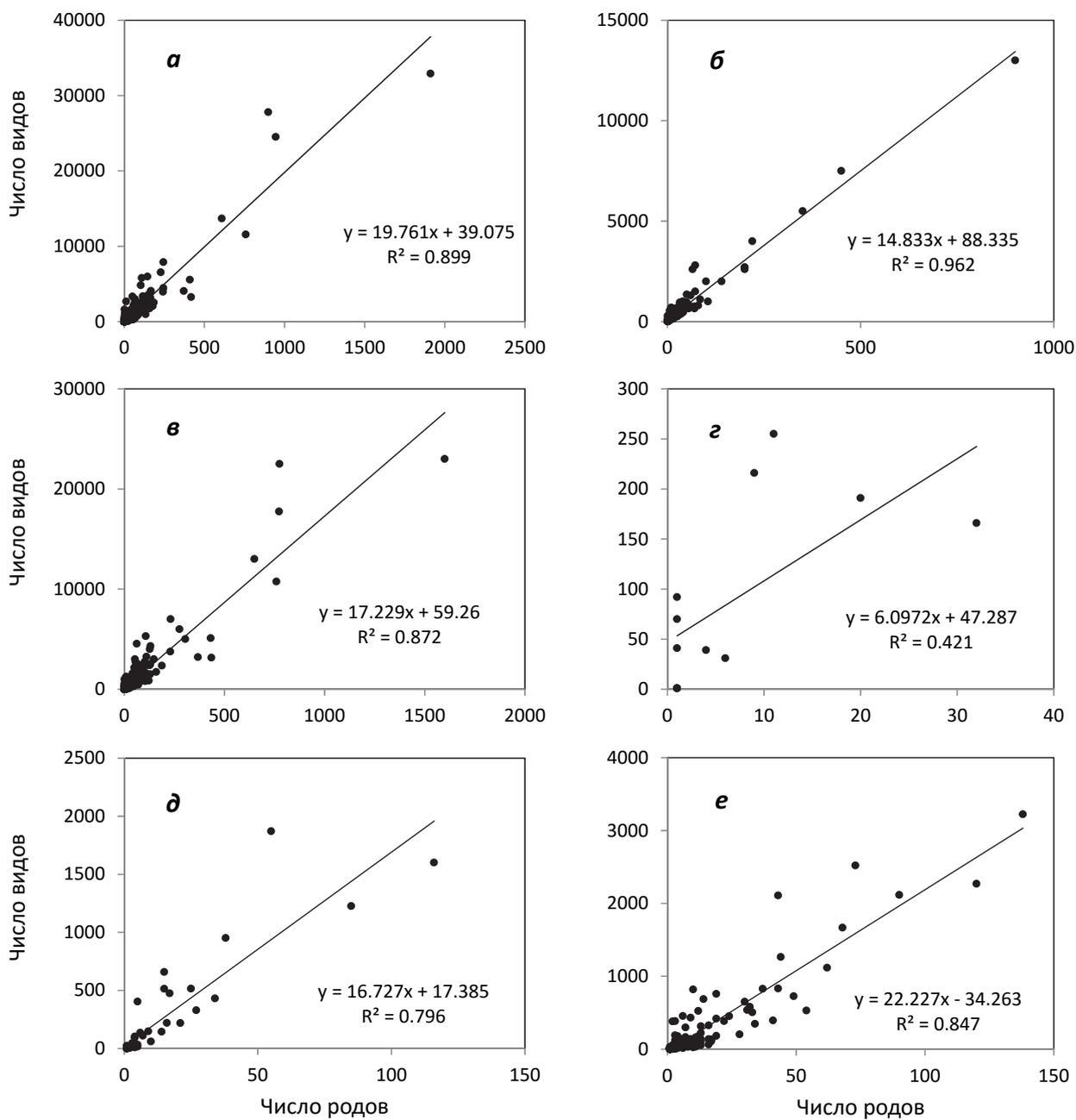

Рис. 3. Соотношение числа родов и видов в семействах покрытосеменных (а, б, в), голосеменных (г), птеридофитов (д) и бриофитов (е). По данным 'The Plant List' (а, г, д, е), J.C. Willis (1919; система А. Энглера) (б), А.Л. Тахтаджяна (Takhtajan, 2009) (в).

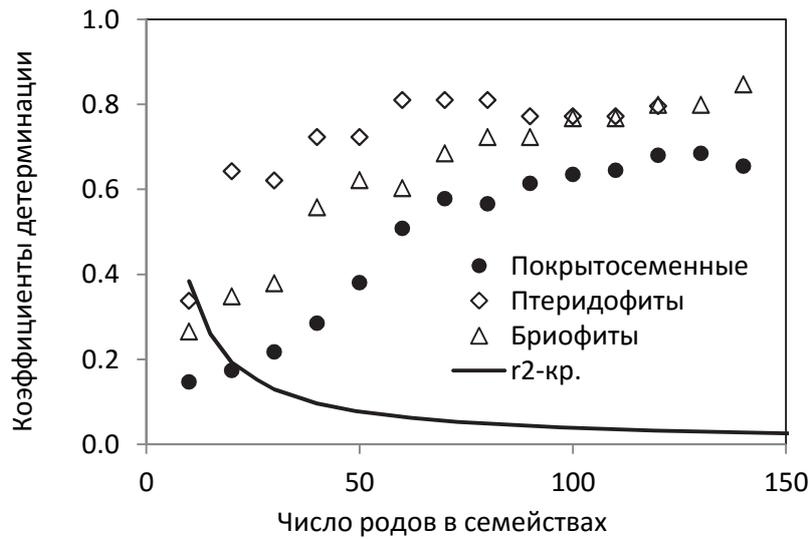

Рис. 4. Коррелированность числа родов и видов в семействах покрытосеменных, птеридофитов и бриофитов в зависимости от объема семейств (объяснения в тексте). $r^2$-кр. - кривая критических значений, ниже которой коэффициенты детерминации статистически не значимы.

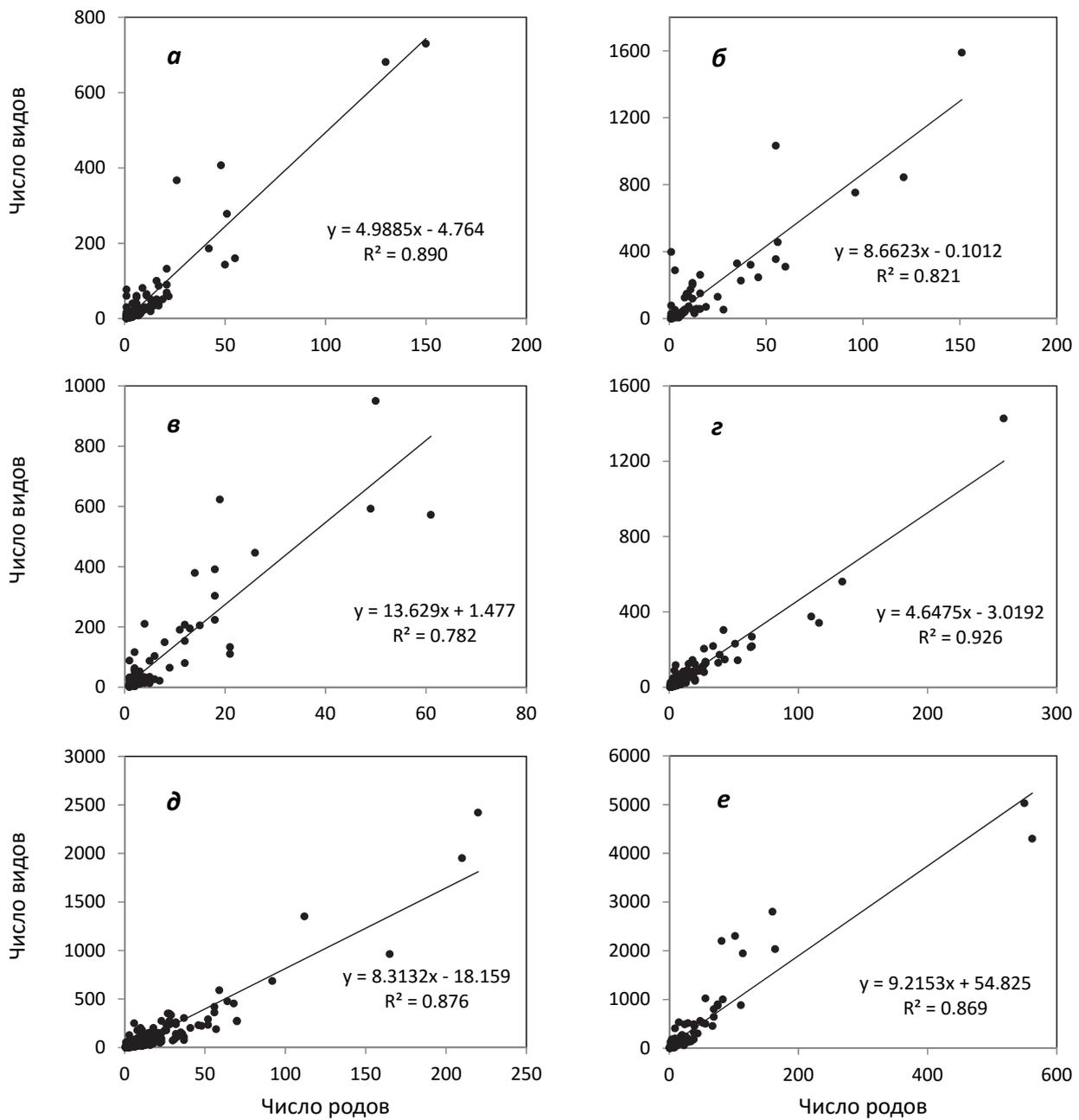

Рис. 5. Соотношение числа родов и видов в семействах млекопитающих (Wilson, Reeder, 2005) (а), рептилий (Uetz, 2010) (б), амфибий (AmphibiaWeb, 2015) (в), птиц (Bock, Farrand, 1980) (г), рыб (Нельсон, 2009) (д), пауков (Jocqué, Dippenaar-Schoeman, 2006) (е).

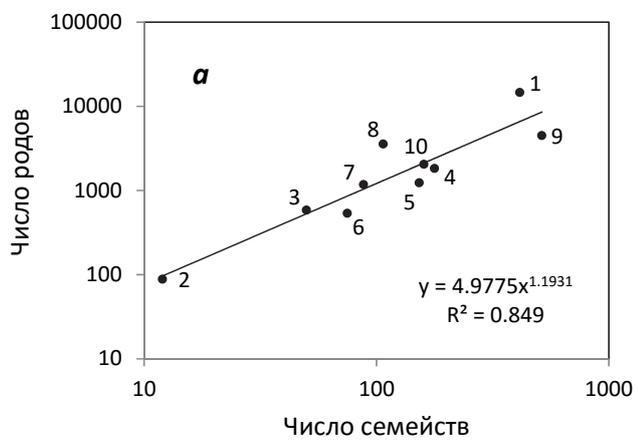 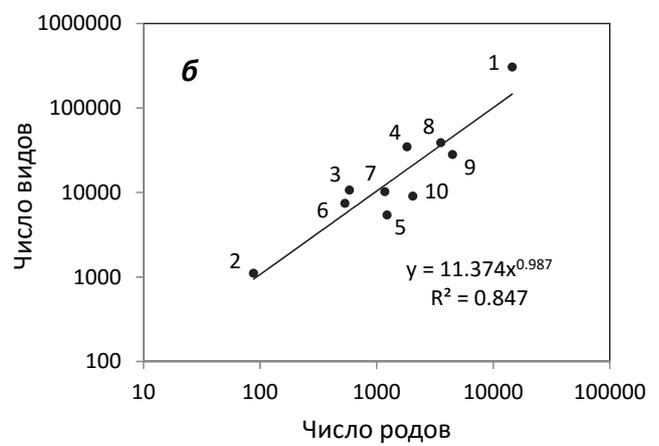

Рис. 6. Соотношения объемов таксонов разных рангов в группах растений и животных. 1 - покрытосеменные, 2 - голосеменные, 3 - птеридофиты, 4 - бриофиты, 5 - млекопитающие, 6 - амфибии, 7 - рептилии, 8 - пауки, 9 - рыбы, 10 - птицы. Оси абсцисс и ординат показаны в логарифмическом масштабе.

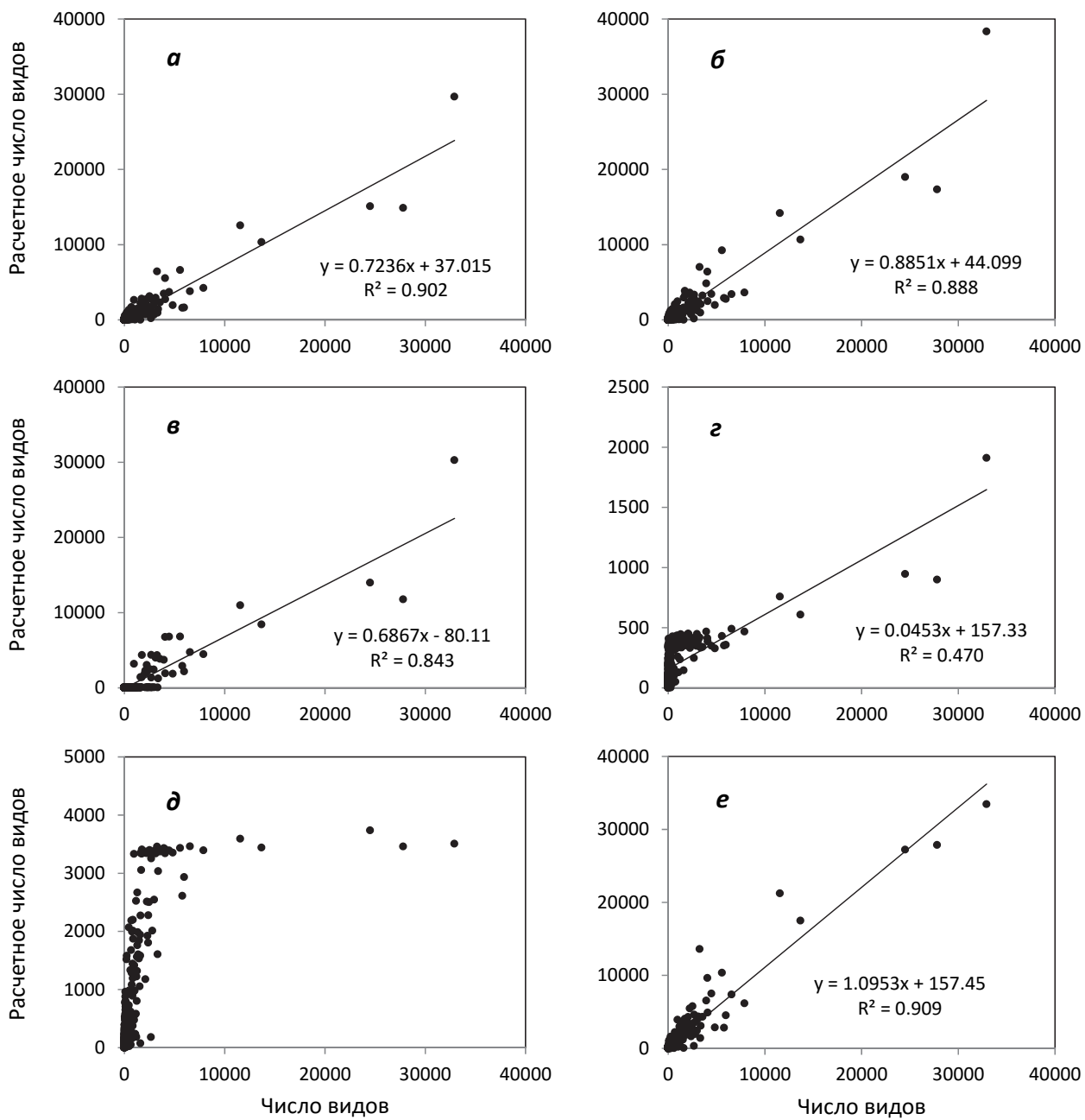

Рис. 7. Соотношение реального и расчетного числа видов по результатам работы программа 'Divergence 3.0' (условия выполнения программы для этого рисунка даны в табл. 2).

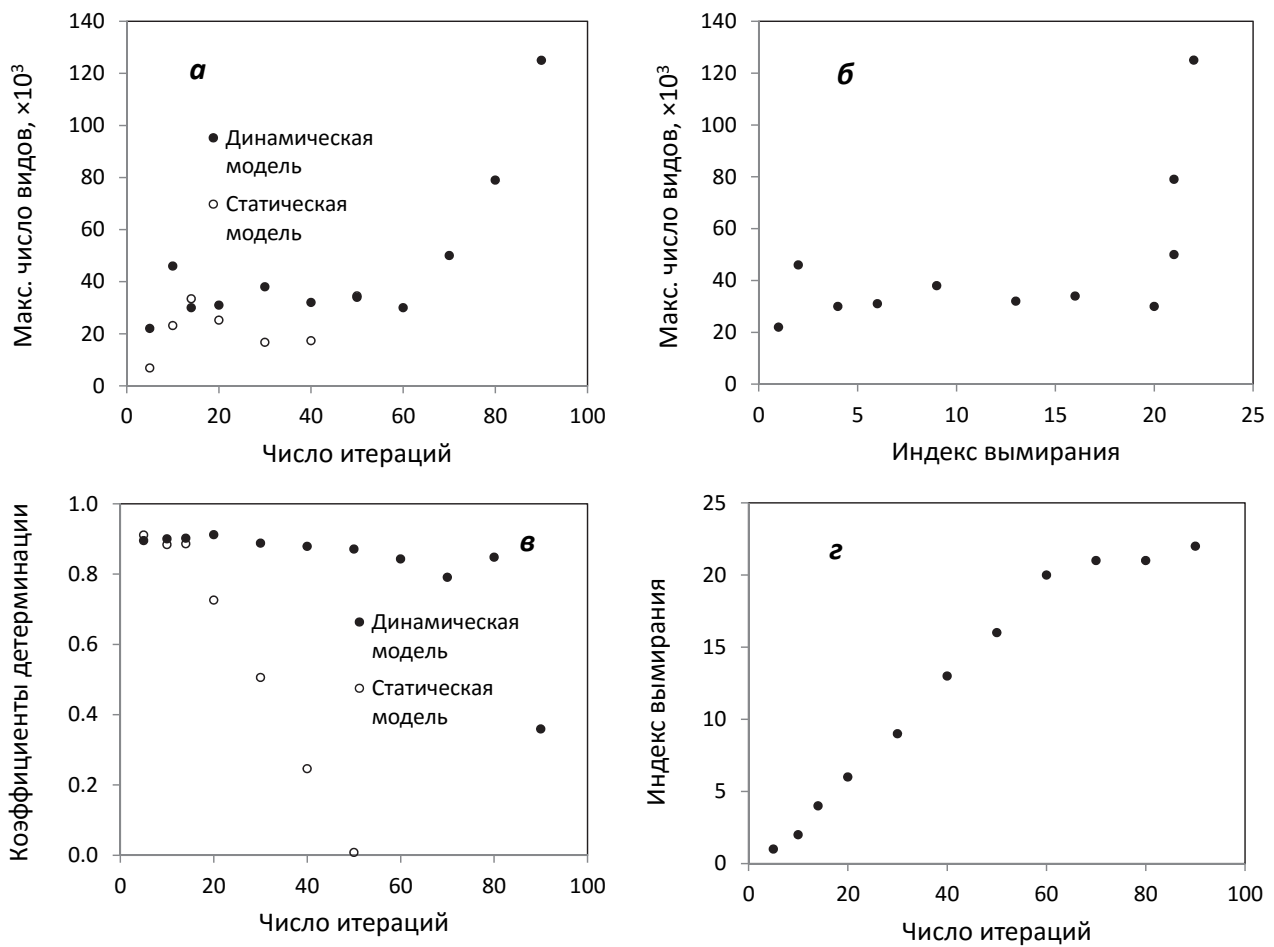

Рис. 8. Анализ результатов работы программы 'Divergence 3.0'. а, б - зависимость максимального расчетного числа видов от числа итераций (а) и индексов вымирания (б); в - зависимость коррелированности между реальным и расчетным числом видов от число итераций в случаях статического и динамического факторов вымирания; г - соотношение между числом итераций и индексами вымирания.

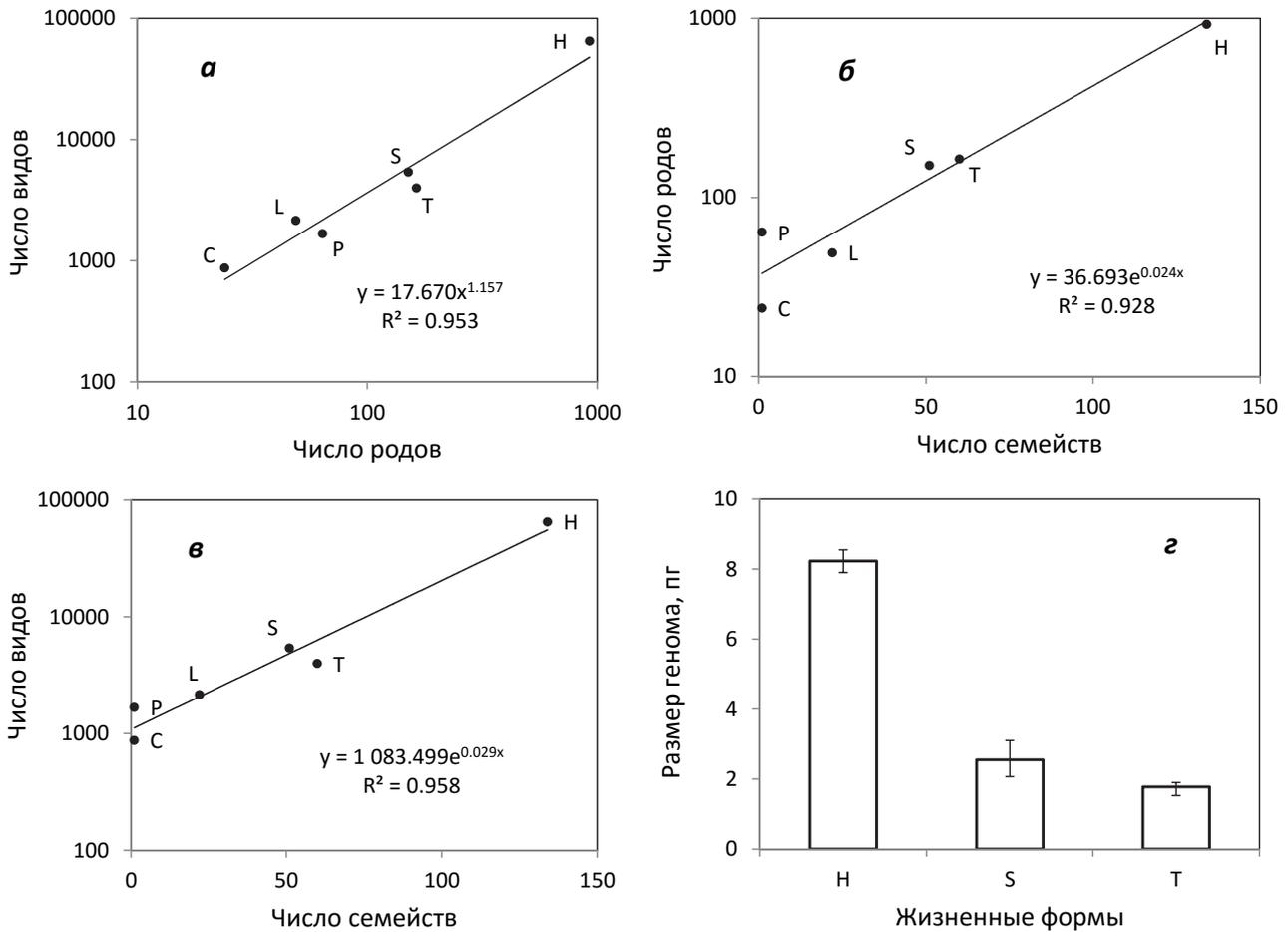

Рис. 9. Соотношение объемов таксонов разных рангов после группировки родов покрытосеменных по жизненным формам (а, б, в; оси ординат в логарифмическом масштабе); размер генома (медианы, пикограммы) у разных жизненных форм покрытосеменных (г, вертикальные планки погрешностей - доверительные интервалы на уровне значимости 0.05). H - травы, S - кустарники, T - деревья, P - пальмы, L - лианы, C - представители сем. Cactaceae.

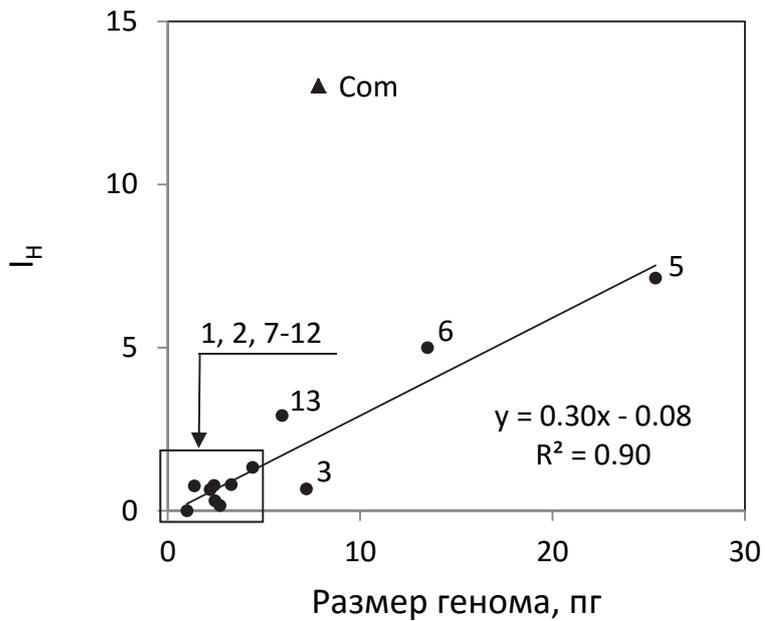

Рис. 10. Соотношение размера генома (медианы) и индекса травянистости ($I_H$) для неформальных групп системы APG III (см. рис. 1) (1 - ANITA grade, 2 - Magnoliids, 3 - basal lineages of Magnoliids, 5 - non Commelinids Monocots, 6 - basal lineages of Eudicots, 7 - basal lineages of Core Eudicots, 8 - Fabids, 9 - Malvids, 10 - basal lineages of Rosids, 11 - basal lineages of Asterids, 12 - Lamiids, 13 - Campanulids). Точка для Commelinids (Com) не включена в расчеты.

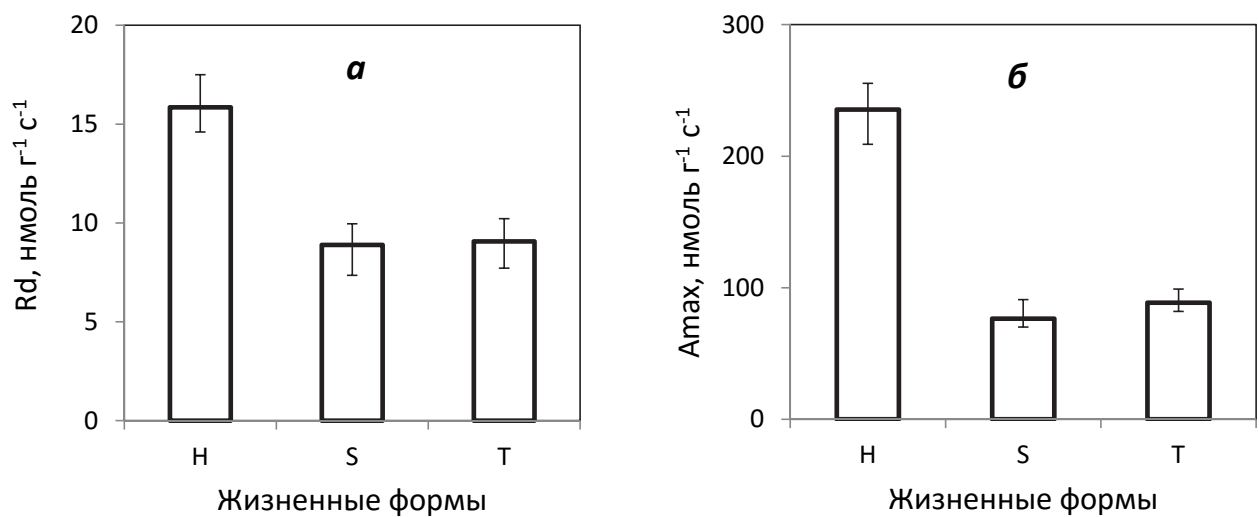

Рис. 11. Темновое дыхание (а) и потенциальный фотосинтез (б) (медианы, вертикальные планки погрешностей - доверительные интервалы на уровне значимости 0.05) у разных жизненных форм по результатам обработки базы данных GLOPNET (Wright et al., 2004).

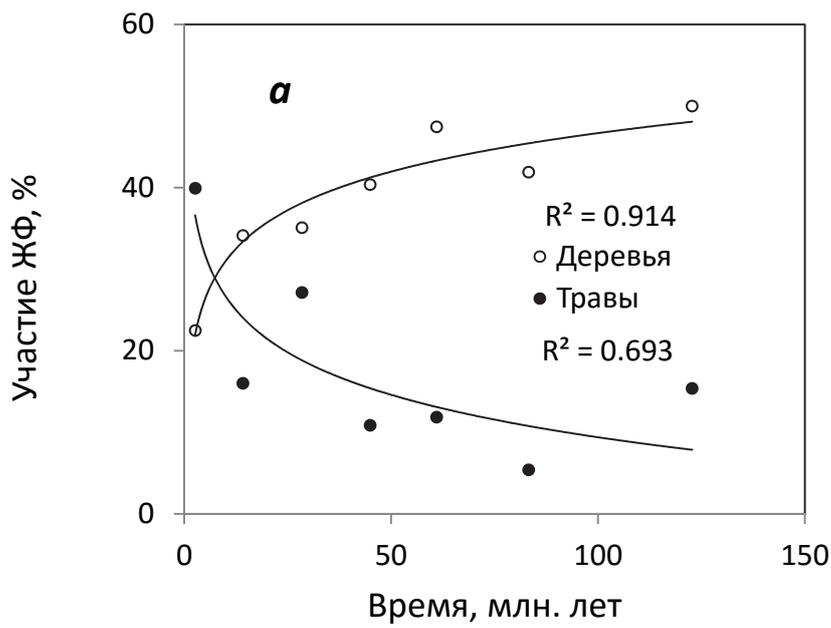
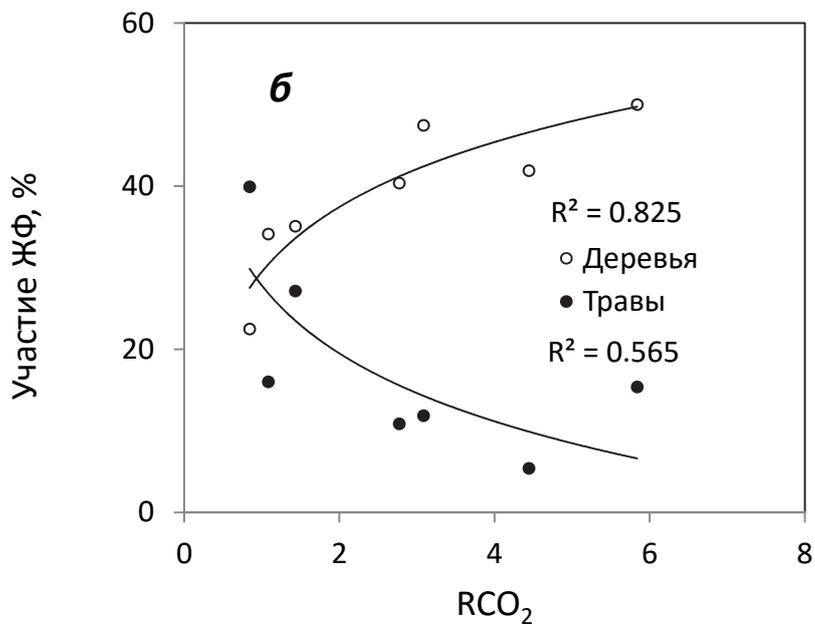

Рис. 12. Участие жизненных форм (ЖФ) в родах покрытосеменных, появившихся впервые в палеонтологической летописи в эпохи мела-кайнозоя (а) и на фоне изменений в это же время относительного (к доиндустриальному уровню) содержания углекислого газа в атмосфере (Berner, 2006) (б).

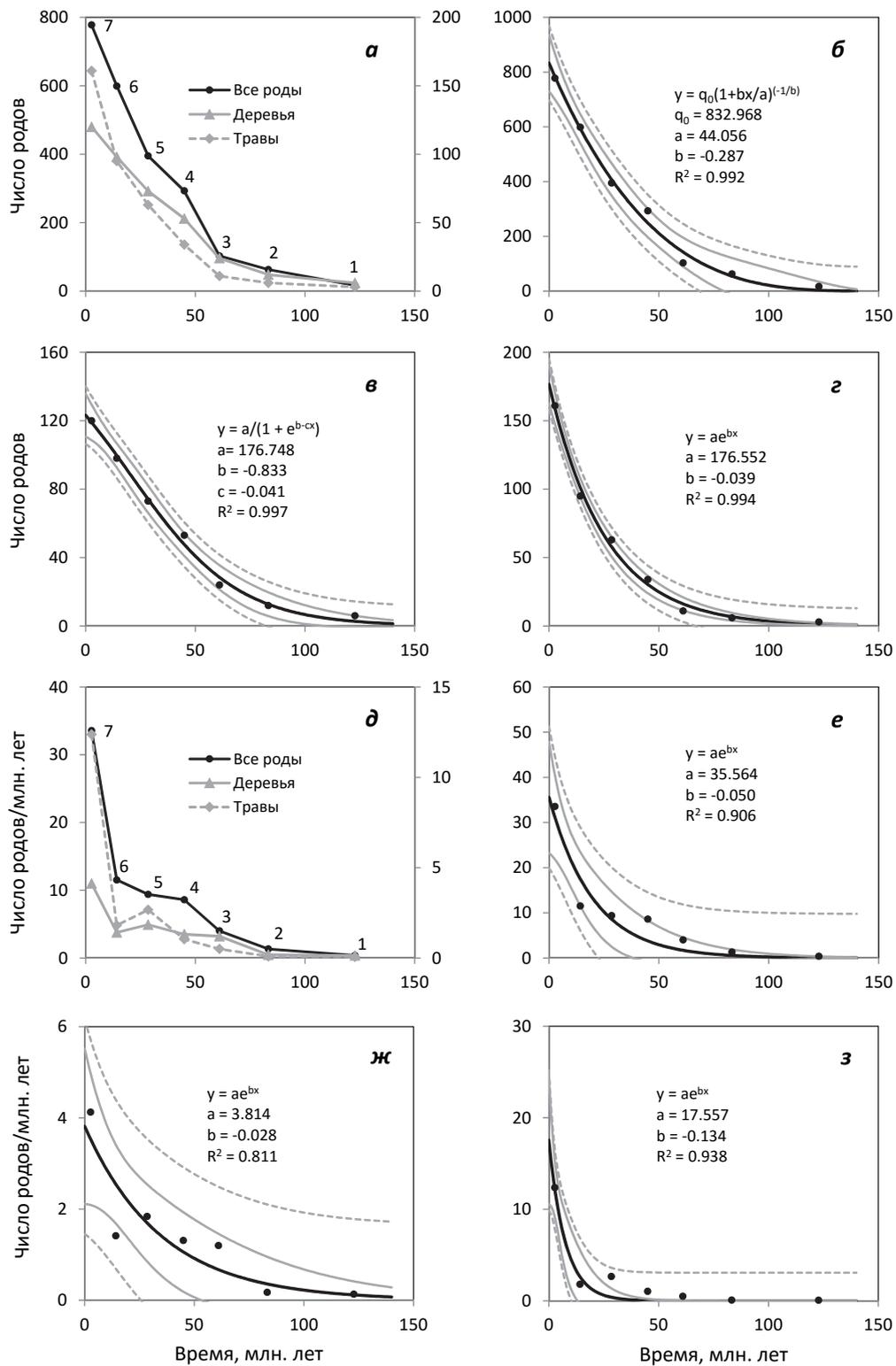

Рис. 13. Изменения числа родов покрытосеменных, появившихся в палеонтологической летописи в эпохи мела-кайнозоя: а- куммулятивные кривые; б, в, г - аппроксимация куммулятивных кривых гиперболой (б, все роды), сигмоидной кривой (в, деревья), экспонентой (г, травы); д - абсолютные скорости диверсификации; е, ж, з - аппроксимация этих скоростей (е - все роды, ж - деревья, з - травы) экспонентами. Светлые сплошные кривые - границы доверительных интервалов, светлые прерывистые кривые - границы интервалов предсказания. Оси ординат (а, д) слева - для всех родов, справа - для деревьев и трав. 1- ранний мел; 2 - поздний мел; 3 - палеоцен; 4 - эоцен; 5 - олигоцен; 6 - миоцен; 7 - плиоцен, плейстоцен, голоцен.

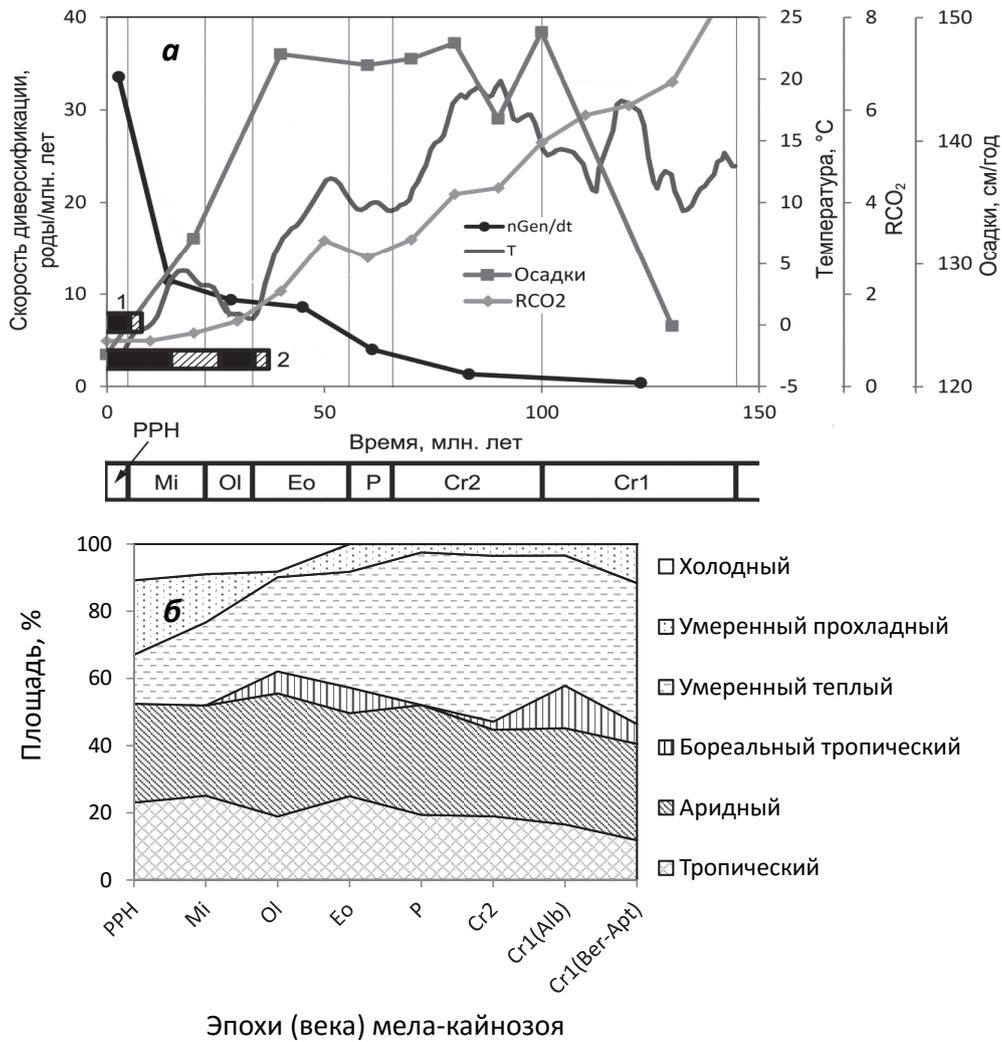

Рис. 14. Климатические условия мела-кайнозоя. а - изменения абсолютной скорости диверсификации (nGen/dt) покрытосеменных на фоне климатических изменений мела-кайнозоя: континентальных осадков (Gibbs et al., 1999); глобальной температуры (T) морской воды, рассчитанной по соотношению изотопов кислорода в раковинах бентосных и планктонных фораминифер (по формуле Lear et al., 2000) по результатам обработки базы данных (см. Veizer et al., 1999); периодов частичного (штриховка) и полного (заливка) оледенения северного (1) и южного (2) полюсов (по: Zachos et al., 2001); относительного содержания углекислого газа в атмосфере ($RCO_2$) по данным модели GEOCARB III (Berner, Kothavala, 2001), континентальных осадков (Gibbs et al., 1999). б - площади территорий, занятых климатическими поясами (в процентах от общей площади суши в соответствующую эпоху) (по картам Boucot et al., 2013; береговые линии - по GPlates ver. 1.5). Cr1(Ber-Apt) - ранний мел (берриас-апт); Cr1(Alb) - ранний мел (альб); Cr2 - поздний мел (кампан); P - палеоцен; Eo - эоцен; Ol - олигоцен; Mi - миоцен, PPH - плиоцен, плейстоцен, голоцен.

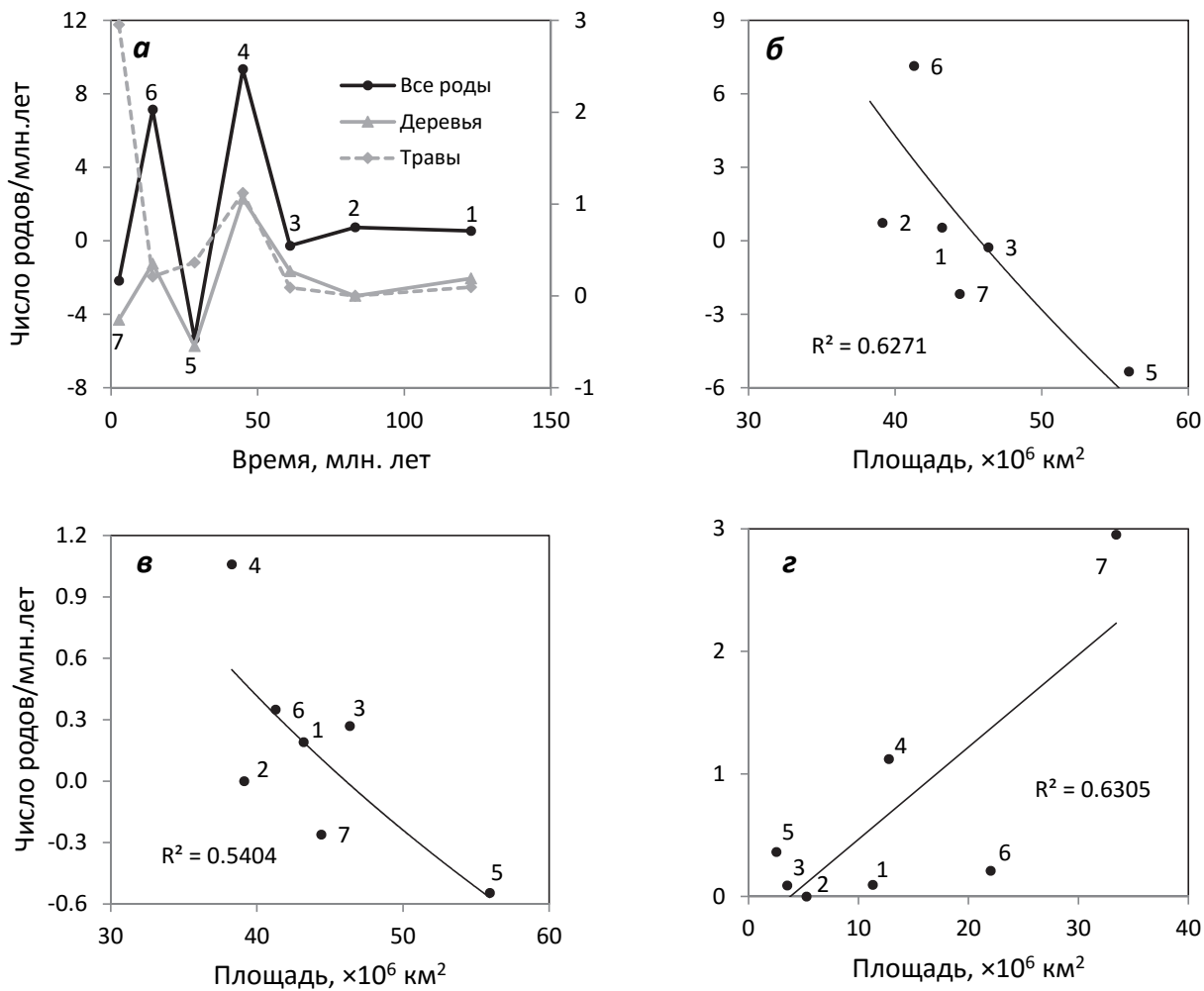

Рис. 15. Относительные скорости диверсификации (а) и их соотношение для всех родов (б), деревьев (в) и трав (г) с площадями (по картам Boucot et al., 2013) аридных (б, в) и умеренно прохладных (г) территорий в эпохи мела-кайнозоя : 1- ранний мел; 2 - поздний мел; 3 - палеоцен; 4 - эоцен; 5 - олигоцен; 6 - миоцен; 7 - плиоцен, плейстоцен, голоцен.

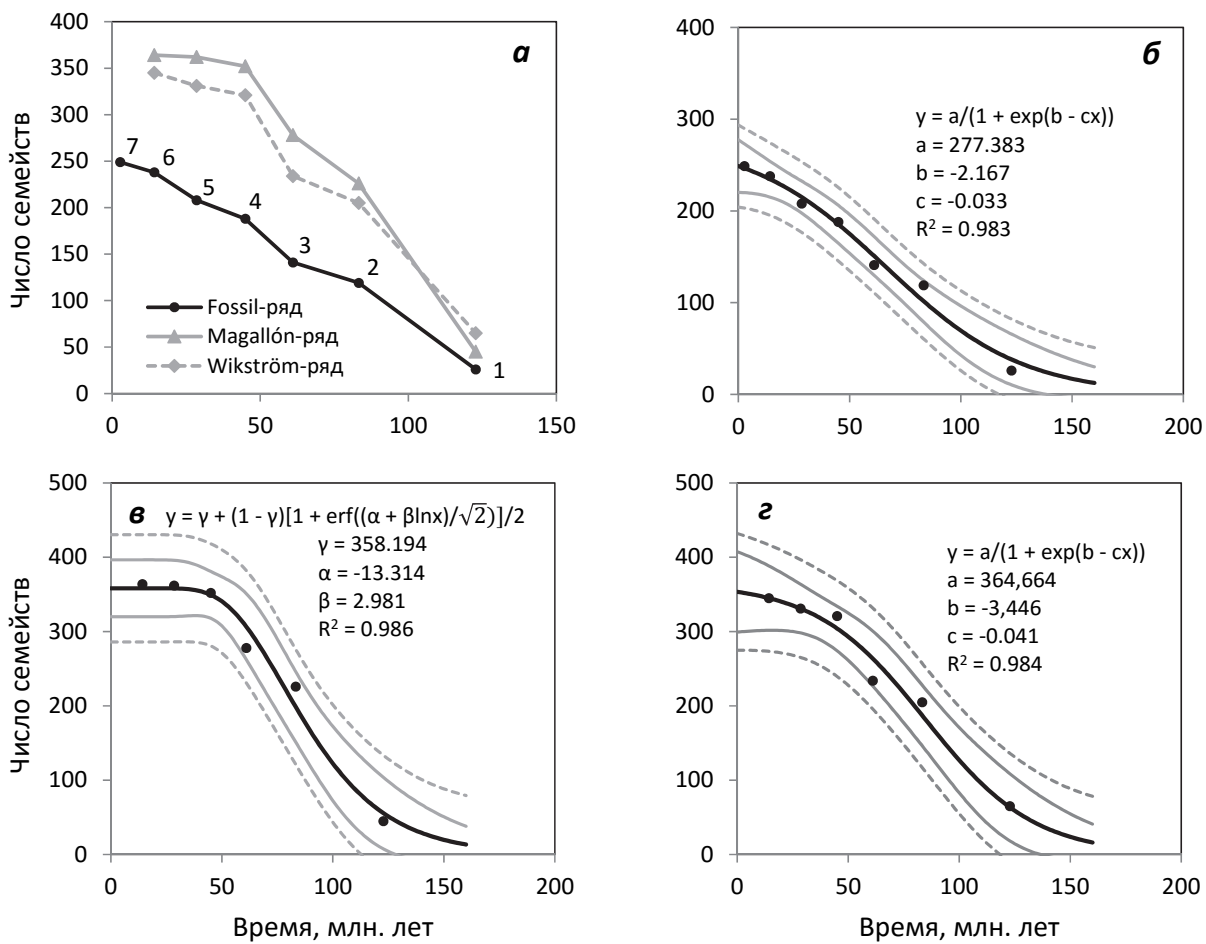

Рис. 16. Изменения числа семейств покрытосеменных, появившихся в палеонтологической летописи в эпохи мела-кайнозоя: а- куммулятивные кривые; б, в, г - аппроксимация куммулятивных кривых (б - Fossil-ряд, в - Magallón-ряд, г - Wikström-ряд) сигмоидными функциями. Светлые сплошные кривые - границы доверительных интервалов; светлые прерывистые кривые - границы интервалов предсказания. 1- ранний мел; 2 - поздний мел; 3 - палеоцен; 4 - эоцен; 5 - олигоцен; 6 - миоцен; 7 - плиоцен, плейстоцен, голоцен.

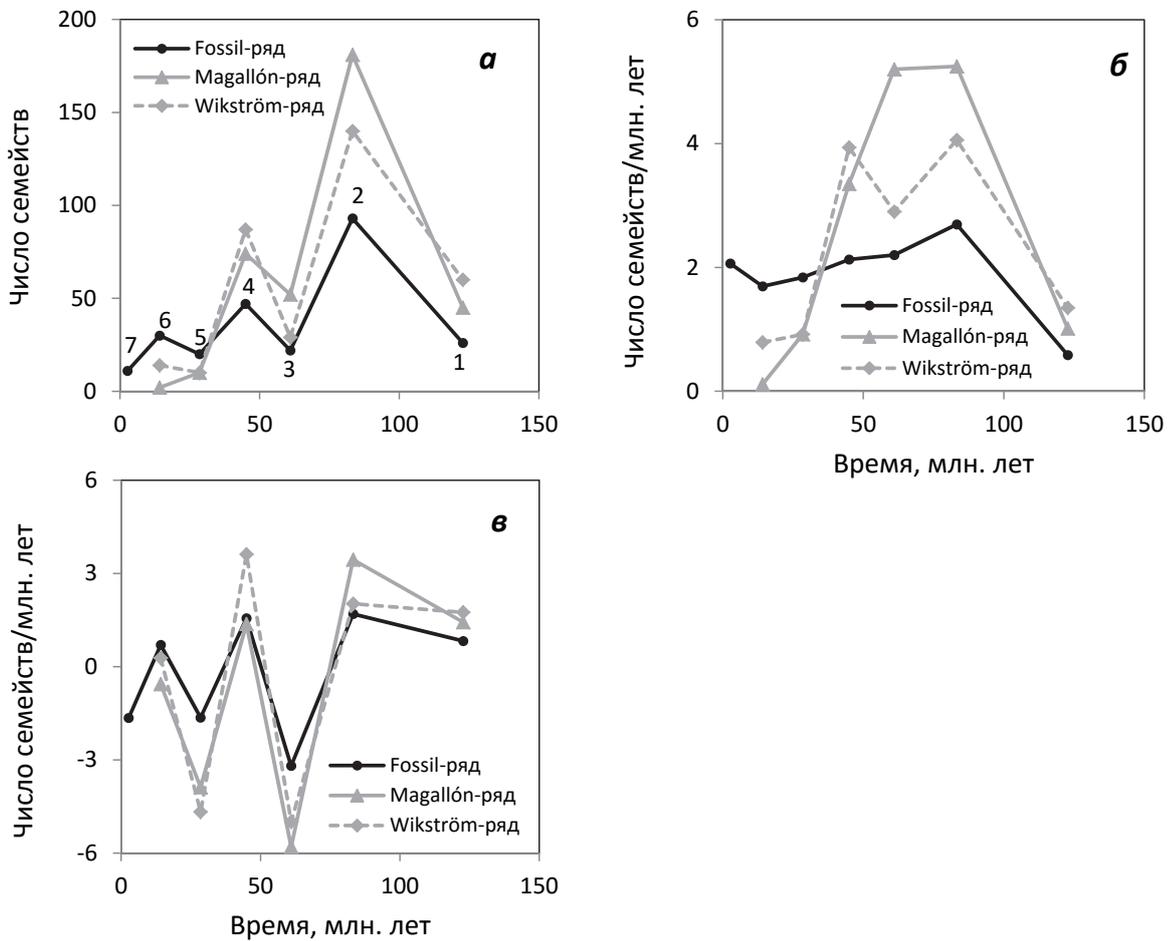

Рис. 17. Изменения числа семейств покрытосеменных, появившихся в палеонтологической летописи в эпохи мела-кайнозоя: а - количество новых семейств, зафиксированных в каждую из эпох; б - то же количество, отнесенное к продолжительности эпох; в - относительная скорость диверсификации семейств покрытосеменных ($\varepsilon = 0$). 1- ранний мел; 2 - поздний мел; 3 - палеоцен; 4 - эоцен; 5 - олигоцен; 6 - миоцен; 7 - плиоцен, плейстоцен, голоцен.

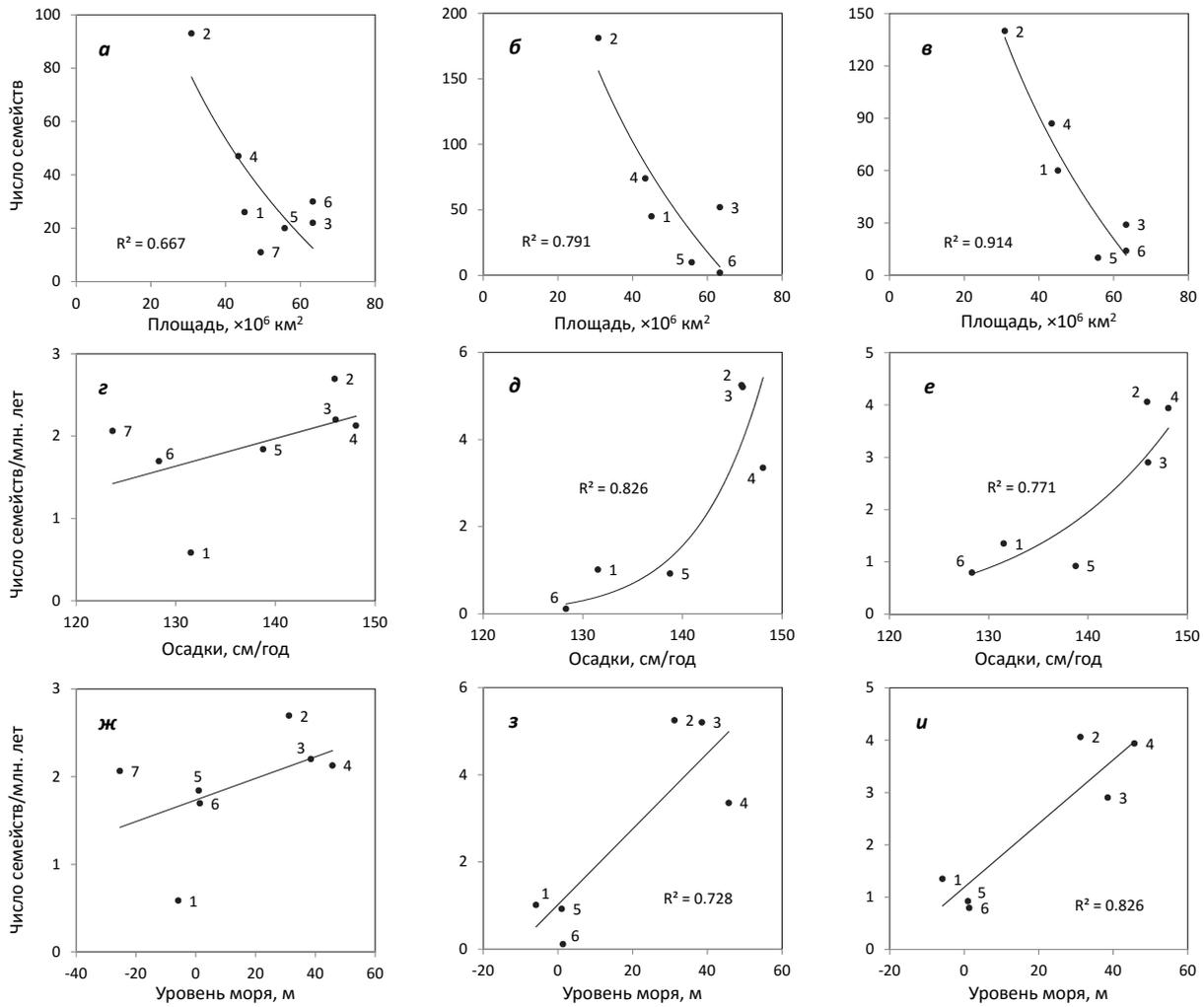

Рис. 18. Соотношение количества семейств, появившихся в эпохи мела-кайнозоя, с площадями аридных территорий (по картам Boucot et al., 2013): а - Fossil-ряд, б - Magallón-ряд, в - Wikström-ряд. Соотношение абсолютных скоростей диверсификации семейств покрытосеменных в эпохи мела-кайнозоя с континентальными осадками (Gibbs et al., 1999) (г, д, е) и уровнем моря (относительно современного) (Miller et al., 2005) (ж, з, и): г, ж - Fossil-ряд; д, з - Magallón-ряд; е, и - Wikström-ряд. 1- ранний мел; 2 - поздний мел; 3 - палеоцен; 4 - эоцен; 5 - олигоцен; 6 - миоцен; 7 - плиоцен, плейстоцен, голоцен.